\documentclass[sigplan,screen]{acmart}

\usepackage{graphicx,subcaption}
\usepackage{subcaption}
\usepackage{color}
\usepackage{enumitem}
\usepackage{wrapfig}
\usepackage{appendix}
\usepackage{xspace}
\usepackage{algcompatible}
\usepackage{algorithm}
\usepackage{algpseudocode}
\usepackage{listings}
\usepackage{verbatim}
\usepackage{multicol}
\usepackage{float}
\usepackage{caption}
\usepackage[utf8]{inputenc}             
\usepackage[T1]{fontenc}                
\usepackage{hyperref}
\usepackage{booktabs}                  
\usepackage{amsfonts}                  
\usepackage{nicefrac}                  
\usepackage{microtype}                 
\usepackage{listings}
\usepackage{xcolor}
\usepackage[disable]{todonotes}
\usepackage{pifont}
\usepackage{subfiles}
\usepackage{multirow}
\usepackage{colortbl}

\graphicspath{ {figs/} }
\definecolor{light-gray}{gray}{0.80}
\usepackage{titlesec}
\titlespacing*{\section}
{0pt}{1ex}{0.6ex}
\titlespacing*{\subsection}
{0.2pt}{0.5ex}{0.2ex}
\titlespacing*{\subsubsection}
{0pt}{0.2ex}{0.5ex}

\AtBeginDocument{%
  \providecommand\BibTeX{{%
    \normalfont B\kern-0.5em{\scshape i\kern-0.25em b}\kern-0.8em\TeX}}}

\setcopyright{rightsretained} 
\copyrightyear{2021}
\acmYear{2021}
\setcopyright{rightsretained}
\acmConference[PPoPP '21]{26th ACM SIGPLAN Symposium on Principles
and Practice of Parallel Programming}{February 27-March 3, 2021}{Virtual
Event, Republic of Korea}
\acmBooktitle{26th ACM SIGPLAN Symposium on Principles and Practice
of Parallel Programming (PPoPP '21), February 27-March 3, 2021, Virtual
Event, Republic of Korea}\acmDOI{10.1145/3437801.3441625}
\acmISBN{978-1-4503-8294-6/21/02}

\newcommand{\punt}[1]{}
\newcommand{\cmnt}[1]{}

\definecolor{xxxcolor}{rgb}{0.8,0,0}

\newcommand{\func}[1]{\texttt{#1}}

\newtheorem{theorem}{Theorem}
\newtheorem{lemma}[theorem]{Lemma}
\newtheorem{corollary}[theorem]{Corollary}

\newtheorem{property}[theorem]{Property}

\newtheorem{requirement}[theorem]{Requirement}
\newcounter{history}

\newtheorem{assumption}[theorem]{Assumption}

\newcommand{\secref}[1]{Section~\ref{sec:#1}}
\newcommand{\figref}[1]{Figure~\ref{fig:#1}}
\newcommand{\tabref}[1]{Table~\ref{tab:#1}}

\newcommand{\lemref}[1]{Lemma~\ref{lem:#1}}

\newcommand{\propref}[1]{Property~\ref{prop:#1}}

\newcommand{\asmref}[1]{Assumption~\ref{asm:#1}}

\newcommand{\Figref}[1]{Figure~\ref{fig:#1}}

\newcommand{\reqref}[1]{Requirement~\ref{req:#1}}

\newcommand{\Lineref}[1]{Line~\ref{lin:#1}}
\newcommand{\lineref}[1]{line~\ref{lin:#1}}
\newcommand{\algoref}[1]{Algorithm~\ref{algo:#1}}

\newcommand{\ignore}[1]{}
\newcommand{\myparagraph}[1]{\noindent\textbf{#1}}

\newcommand{\nz}{\emph{restartable}\xspace}

\newcommand{\rdp}{\emph{$\Phi_{read}$}\xspace}
\newcommand{\wtp}{\emph{$\Phi_{write}$}\xspace}

\newcommand{\rd}{\emph{reader}\xspace}
\newcommand{\wt}{\emph{writer}\xspace}
\newcommand{\rl}{\emph{reclaimer}\xspace}

\newcommand{\nbr}{\emph{NBR}\xspace}
\newcommand{\nbrp}{\emph{NBR+}\xspace}

\newcommand{\ds}{\emph{data-structure}\xspace}

\newcommand{\lb}{\emph{limboBag}\xspace}
\newcommand{\hw}{\emph{HiWatermark}\xspace}
\newcommand{\lw}{\emph{LoWatermark}\xspace}
\newcommand{\smr}{\emph{safe memory reclamation}\xspace}

\newcommand{\tid}[1]{$T_#1$}

\newcommand{\rcu}{\emph{RCU}\xspace}
\newcommand{\qsbr}{\emph{QSBR}\xspace}
\newcommand{\debra}{\emph{DEBRA}\xspace}
\newcommand{\ibr}{\emph{IBR}\xspace}
\newcommand{\hp}{\emph{HP}\xspace}

\newcommand{\rgp}{\emph{RGP}\xspace}

\begin{document}

\title{NBR: Neutralization Based Reclamation}


\author{Ajay Singh}
\affiliation{%
  \institution{University of Waterloo}
  \country{Canada}}
\email{ajay.singh1@uwaterloo.ca}

\author{Trevor Brown}
\affiliation{%
  \institution{University of Waterloo}
  \country{Canada}}
\email{trevor.brown@uwaterloo.ca}

\author{Ali Mashtizadeh}
\affiliation{%
  \institution{University of Waterloo}
  \country{Canada}}
\email{mashti@uwaterloo.ca}


\begin{abstract}
\textit{Safe memory reclamation} (SMR) algorithms suffer from a trade-off between bounding unreclaimed memory and the speed of reclamation.
Hazard pointer (HP) based algorithms bound unreclaimed memory at all times, but tend to be slower than other approaches.
Epoch based reclamation (EBR) algorithms are faster, but do not bound memory reclamation.
Other algorithms follow hybrid approaches, requiring special compiler or hardware support, changes to record layouts, and/or extensive code changes.
Not all SMR algorithms can be used to reclaim memory for all data structures.

We propose a new \textit{neutralization} based reclamation (NBR) algorithm that is often faster than the best known EBR algorithms and achieves bounded unreclaimed memory.
It is non-blocking when used with a non-blocking operating system (OS) kernel, and only requires atomic read, write and CAS.
NBR is straightforward to use with many different data structures, and in most cases, requires similar reasoning and programmer effort to two-phased locking.
\nbr is implemented using OS signals and a lightweight handshaking mechanism between participating threads to determine when it is safe to reclaim a record.
Experiments on a lock-based binary search tree and a lazy linked list show that \nbr significantly outperforms many state of the art reclamation algorithms.
In the tree, NBR is faster than next best algorithm, DEBRA, by up to 38\% and HP by up to 17\%. And, in the list, NBR is 15\% and 243\% faster than DEBRA and HP, respectively.  
\end{abstract}



\keywords{safe memory reclamation, synchronization and concurrency control, concurrent data structures, algorithms.}


\maketitle

\lstdefinestyle{ajstyle}{ %
	language=C++,
	numbers=left,                    
	morekeywords={*, startOp,...},   
	breaklines=true,                 
	frame=single,
belowcaptionskip=1\baselineskip,
showstringspaces=false,
basicstyle=\scriptsize\ttfamily,
keywordstyle=\bfseries\color{green!40!black},
commentstyle=\itshape\color{purple!40!black},
identifierstyle=\color{blue},
stringstyle=\color{orange},	
}

\lstdefinestyle{ajstyleds}{ %
	language=C++,
	morekeywords={*, startOp,...},   
	breaklines=true,                 
	frame=single,
belowcaptionskip=1\baselineskip,
xleftmargin=\parindent,	
showstringspaces=false,
basicstyle=\scriptsize\ttfamily,
keywordstyle=\bfseries\color{green!40!black},
commentstyle=\itshape\color{purple!40!black},
identifierstyle=\color{blue},
stringstyle=\color{orange},	
numberstyle=\scriptsize,
}
\lstset{escapechar=|,style=ajstyle}

\section{Introduction}
\label{sec:intro}

%

Fundamentally, {\em safe memory reclamation} (SMR) is about answering the question: When is it safe to free a record? 
Unlike garbage collection, which is automatic, SMR requires a program to invoke a \textit{retire} operation on each record at some point after it becomes \textit{garbage} (i.e., is \textit{unlinked} from the data structure).
The task of an SMR algorithm is to eventually \textit{free} an unlinked record once no thread holds a pointer to it ~\cite{brown2015reclaiming, michael2004hazard, braginsky2013drop}.

The challenge of SMR in concurrent data structures comes from \func{use-after-free} 
conflicts between threads, where one thread accesses a record that is concurrently freed by another.
For example, consider a lazy-list where one thread is searching and another is deleting.
The first thread obtains a reference to a record and stores it in a local variable.
The other thread unlinks and frees.
At this point the first thread's reference is no longer safe as the record it points to has been freed.

Researchers have developed a rich variety of SMR algorithms, with a diverse spectrum of desirable properties, idiosyncrasies and limitations.
After experimenting with SMR algorithms and observing the state of art ~\cite{brown2015reclaiming,michael2004hazard,detlefs2002lock,gidenstam2008efficient,hart2007performance,cohen2015automatic,balmau2016fast,alistarh2018threadscan,cohen2018every,wen2018interval,dice2016fast,herlihy2005nonblocking,blelloch2020concurrent,ramalhete2017brief,cohen2015efficient, alistarh2014stacktrack, dragojevic2011power, alistarh2017forkscan, braginsky2013drop, nikolaev2019hyaline}, we identified the following set of desirable properties. 
[P1]~\textit{Performance}: reclamation operations should ideally offer both low latency and high throughput.
[P2]~\textit{Bounded Garbage}: The number of records that are unlinked but not yet reclaimed should be bounded, even if threads experience halting failures or long delays.
[P3]~\textit{Usability}: Intrusive changes to code, data, and the build environment, should be minimized.
[P4] \textit{Consistency}: Performance should not be drastically affected by changes in the workload (e.g., when shifting between read-intensive and update-intensive workloads).  Additionally, there should be minimal performance degradation when the system is \textit{oversubscribed} (with more threads than cores). 
[P5]~\textit{Applicability}: The algorithm should be usable with as many data structures as possible.

To set the stage for our contribution, we must first discuss other approaches.
We broadly categorize existing work into: 
hazard pointer-based reclamation (HPBR), quiescent state-based reclamation (QSBR), epoch-based reclamation (EBR), reference counting based reclamation (RCBR), and hybrid algorithms that combine the aforementioned approaches ~\cite{hart2007performance}.
In general, QSBR and EBR are fast but do not bound garbage, HPBR has bounded garbage but is not fast, and RCBR is neither fast nor does it bound garbage (in case retired nodes have pointer cycles~\cite{detlefs2002lock}).
Hybrid approaches have generally focused on achieving P1 and P2 simultaneously, usually by combining EBR (for its speed) with some variant of HPBR (to bound garbage), with varying levels of success.

The hybrid algorithm that most closely resembles our approach is DEBRA+~\cite{brown2015reclaiming}, a variant of EBR (with a restricted form of HPBR) that is designed for lock-free data structures. 
DEBRA+ is fast, and it achieves bounded garbage via a \textit{neutralizing} mechanism based on POSIX signals and data structure specific recovery code.
A thread whose reclamation is delayed by a slow thread will send a \textit{neutralizing signal} to the slow thread.
Upon receipt of a neutralizing signal, a thread executes its recovery code and then restarts its data structure operation, 
allowing reclamation to continue, ultimately guaranteeing a bound on the number of unreclaimed records. 
However, this bound on garbage comes at the cost of both usability and applicability, as users need to write data structure specific recovery code that is not always straightforward, or even possible.
Moreover, it is not clear how DEBRA+ could be used for lock-based data structures, since neutralizing a thread that holds a lock could cause deadlock.

\myparagraph{Contribution}:
Existing SMR algorithms all have significant shortcomings in their attempts at satisfying properties P1 through P5.
This motivated us to propose a new \textit{Neutralization Based Reclamation} algorithm (\nbr) that matches or outperforms existing SMR algorithms [P1], bounds garbage [P2], is simple to use [P3], exhibits consistent performance, even on oversubscribed systems [P4], and is applicable to a large class of data structures, some of which are not supported by popular SMR algorithms [P5].

\nbr's neutralization technique is similar to that of DEBRA+, with a few key differences.
In \nbr, each thread places unlinked objects in a thread-local buffer, and when the buffer's size exceeds a predetermined threshold, the thread sends a neutralizing signal to \textit{all} other threads.
Upon receipt of such a signal, a thread checks whether its current data structure operation has already done any writes to shared memory, and if not, restarts its operation (using the C/C++ procedures \func{sigsetjmp} and \func{siglongjmp}).
Otherwise, it finishes executing its operation.
In contrast, to guarantee bounded garbage in DEBRA+, a thread must restart even if it has already written to shared memory---a design decision that limits DEBRA+'s applicability to specific lock-free data structures, and necessitates data structure specific recovery code.
\nbr does not require any recovery code, and can be used with nearly all structures that DEBRA+ supports and many others structures DEBRA+ does not, including some \textit{lock-based algorithms}, such as a lock-based binary search tree with lock-free searches~\cite{david2015asynchronized} (DGT).

We also present an optimized version of \nbr called \nbrp in which threads send fewer signals, and yet reclaim memory more often.
This is accomplished by allowing threads to infer when memory can be freed simply by passively observing the signals sent in the system. 
Finally, as our experiments show, NBR+ is highly efficient, significantly outperforming the state of the art in SMR in various data structure workloads on a large-scale 4-socket Intel system. 


The rest of the paper is structured as follows.
Related work is surveyed in Section~\ref{sec:related}.
In Section~\ref{sec:model}, we introduce the model.
Section~\ref{sec:nbr} describes our basic algorithm \nbr, and characterizes its applicability.
We describe an optimized version \nbrp in Section~\ref{sec:nbrp}, followed by a brief discussion on \nbr{'s} correctness in \secref{corr}.
Finally, experiments appear in Section~\ref{sec:eval}, followed by conclusions in Section~\ref{sec:conclusion}.


\section{Related Work} \label{sec:related}

Although detailed surveys of \smr already exist in earlier works~\cite{hart2007performance, brown2015reclaiming}, we would like to study 
existing techniques specifically through the lens of the desirable properties defined above.

\myparagraph{RCBR} involves explicitly counting the number of incoming pointers to a record, and typically storing this count alongside the record. 
The inclusion of this metadata in 
records complicates any advanced pointer arithmetic or implicit pointers, and can require changes to record layouts (or the use of a custom allocator) as well as size. 
RCBR typically requires a programmer to invoke a \textit{deref} operation to dereference a pointer (and sometimes to explicitly invoke operations for read, write and CAS)~\cite{detlefs2002lock, herlihy2005nonblocking, blelloch2020concurrent, nikolaev2019hyaline}, adding significant overhead and programmer effort [opposing P1, P3]. 
Programmer intervention is also needed to identify and break pointer \textit{cycles} in garbage records.

\myparagraph{HPBR} incurs significant overhead every time a new record is encountered, as a thread must first \textit{announce} a hazard pointer (HP) to it in a shared location, then issue a memory fence (or use an atomic exchange instruction to announce the hazard pointer) and then check whether the record has already been unlinked~\cite{michael2004hazard, herlihy2005nonblocking, dice2016fast} [opposing P1, P3]. 
If the record has been unlinked, the data structure operation trying to access it must be \textit{restarted} (a data structure specific action).
Correctly dealing with such failure cases can require extensive code changes. 
This may also require the programmer to \textit{reprove} the data structure's progress guarantees~\cite{brown2015reclaiming}.
Additionally, it is not clear how HPs could be used with data structures that allow threads to \textit{traverse pointers in unlinked records}~\cite{brown2015reclaiming}, and there are \textbf{many} examples of such data structures, e.g.,~\cite{heller2005lazy, Brown:2014, afek2014cb, drachsler2014practical, ellen2010non, natarajan2014fast, shafiei2013non, bronson2010practical, fatourou2019persistent, prokopec2012concurrent} [opposing P5].
(In such data structures, a search can potentially pass through many unlinked records, and yet end up back in the data structure, at the appropriate location.)

The latter limitation was addressed by \textit{Beware and Cleanup}---a hybrid of RCBR and HPBR~\cite{gidenstam2008efficient}.
However, this algorithm requires a programmer to write a data structure specific \textit{cleanup} procedure that changes all pointers in an unlinked record to point to \textit{current} records \textit{in the data structure}.
This cleanup code ensures bounded garbage for data structures that allow traversing unlinked records, but the algorithm has higher overhead than either RCBR or HPBR and requires significant programmer effort [opposing P1, P3].

\myparagraph{QSBR and EBR}~\cite{hart2007performance,fraser2004practical,mckenney1998read} both use the observation that, in many data structures, threads do \textit{not} carry pointers obtained in one data structure operation forward for use in a subsequent operation.
QSBR and EBR 
each have a simple interface in which the programmer need only invoke a specific operation at the start and end of a data structure operation.
Unlike the approaches above, QSBR and EBR avoid all per-record and per-access overheads.
A thread can reclaim its garbage records whenever it detects that all other threads have started a new data structure operation (and hence \textit{forgotten} all pointers to said garbage records).
However, in the event that a thread halts or is delayed, the amount of unreclaimed garbage can grow unboundedly 
[opposing P2]. 

DEBRA+ (described above) was introduced by Brown in 2015~\cite{brown2015reclaiming}.
In the same paper, an algorithm called DEBRA was proposed which, to the best of our knowledge, is the fastest EBR algorithm.
DEBRA does not bound the number of unreclaimed records (garbage), but was shown to be faster than DEBRA+.
Note that our experiments show NBR+ often \textit{outperforms} DEBRA.

Since DEBRA, numerous hybrid algorithms offering bounded garbage have appeared, for example, Hyaline (HY)~\cite{nikolaev2019hyaline}, Hazard Eras (HE)~\cite{ramalhete2017brief}, Interval Based Reclamation (IBR)~\cite{wen2018interval}, and Wait Free Eras (WFE)\cite{nikolaev2020universal}. 
All of these algorithms use per-record metadata to 
encode the times at which a record is allocated and unlinked, and the code instrumentation needed is similar to HPs [opposing P3].
HPs require per-record reserve and unreserve calls and fallback code (to restart the operation) in case the reservation fails (because the record is, or might be, unlinked).
It is unclear how HE, IBR and WFE could be used with data structures that allow traversing unlinked records [opposing P5].
As we will see in our experiments, these algorithms also incur non-trivial overhead [opposing P1]. 

Various other algorithms utilize operating system features such as forced context switches~\cite{balmau2016fast}, POSIX signals~\cite{alistarh2017forkscan,alistarh2018threadscan}, and hardware transactional memory~\cite{alistarh2014stacktrack,dragojevic2011power}.
QSense~\cite{balmau2016fast} is a hybrid algorithm that uses QSBR as a \textit{fast} code path, and HPs with forced context switches as a \textit{slow} code path to bound garbage.
However, in the event of long thread delays, reclamation can only proceed on the \textit{slow} path, which is as slow as HPBR.
QSense has been shown to be slower than EBR~\cite{balmau2016fast} [opposing P1].
None of \cite{alistarh2018threadscan, alistarh2014stacktrack, balmau2016fast} can be used with 
data structures that allow threads to traverse unlinked records [opposing P5].
Forkscan (FS)~\cite{alistarh2017forkscan} was succeeded by ThreadScan (TS)~\cite{alistarh2018threadscan}, which addressed this issue, but FS assumes the programmer will not use advanced pointer arithmetic techniques (or implicit pointers) [opposing P3].
Additionally, FS has been shown to be slower than HPs in several workloads~\cite{alistarh2017forkscan} [opposing P1]. 


Optimistic Access (OA) and Automatic Optimistic Access (AOA)~\cite{cohen2015efficient,cohen2015automatic} proposed a particularly interesting approach: they optimistically allow threads to \textit{accesses reclaimed nodes}, and verify \textit{after the fact} that the access was safe.
This requires an assumption that either (a) memory will not be \func{free}d to the OS, or (B) any resulting trap/exception (such as a segmentation fault) will be caught and handled [opposing P3].
Additionally, OA requires the programmer to transform data structures into a \textit{normalized form}~\cite{timnat2014practical} (which is similar to, but not the same as, the form we assume in this paper), and instrument every read/write/CAS [opposing P3].
AOA automates this transformation with compiler support (for data structures that \textit{have} a normalized form).
Unfortunately, it doesn't appear that AOA has been ported to modern compilers.
The need for a normalized form was eliminated in Free Access (FA)~\cite{cohen2018every}, which used a compiler extension to perform automatic instrumentation of writes and blocks of consecutive independent reads.
FA is a general technique that has been shown to have comparable performance to HPBR~\cite{cohen2018every} [opposing P1].
In contrast, our work targets applications that can benefit from the high performance handcrafted SMR. 

\section{Model}
\label{sec:model}


We consider an $n$ thread asynchronous shared memory system.
Threads can perform atomic read, write, compare-and-swap (CAS) and fetch-and-add (FAA).
A data structure consists of a set of records which are accessible from a \textit{root} (e.g., the head of a list).
A \emph{record} can be viewed as a set of fields.
Each record can be in one of five states throughout its lifecycle:
(1) \emph{allocated}: record allocated from heap but not accessible through the root,
(2) \emph{reachable}: the record can be reached by following references from the root,
(3) \emph{unlinked}: is not reachable from (any) root but threads may still have references to it in thread private memory,
(4) \emph{safe}: a record is unlinked and no thread has a reference to it, and
(5) \emph{reclaimed} (or freed): a record is returned to the OS.
In states 3 and 4, a record is \textit{garbage}.

\setlength{\textfloatsep}{5pt}
\begin{figure}[t]
\centering
\resizebox{1\linewidth}{!}{\input{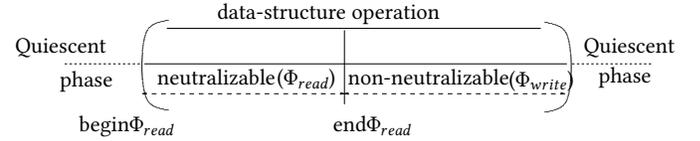}}
\vspace*{-3mm}
\caption{Visualizing the form of a \ds operation for which \nbr can be used. The thread performing this operation can be neutralized in the read-phase (\rdp). However, it cannot be neutralized in the write-phase (\wtp). end\rdp marks the beginning of the operation's \wtp.} 
\label{fig:3phase}
\end{figure}

\section{NBR}
\label{sec:nbr}

\subsection{Assumptions on the data structure}
\label{sec:assumptions}


\nbr requires a data structure's operations to have (or be restructured into) the following form.
This form is needed for the neutralization mechanism wherein an operation that has not yet written to shared memory is forced to restart.
The required form is described as a sequence of \textit{phases} (illustrated in \figref{3phase}), with specific rules in each phase.
Along the way, we 
discuss potential pitfalls for readers unfamiliar with the use of \func{sigsetjmp} and \func{siglongjmp}. 
\\
\\
\noindent \textbf{Phase 0:} preamble. Accesses (reads/writes/CASs) to global variables are permitted. System calls (heap allocation/deallocation, file I/O, network I/O, etc.) are permitted. Access to shared records, for example, nodes of a shared data structure, is \textit{not} permitted. 

\noindent \textbf{Phase 1:} \rdp ({\em read phase}). Reading global variables is permitted and reading shared records is permitted if pointers to them were obtained during this phase (e.g., by traversing a sequence of shared objects by following pointers starting from a global variable---i.e., a \textit{root}). Writes/CASs to shared records, writes/CASs to shared globals, and system calls, are \textit{not} permitted.

To understand the latter restriction, suppose an operation allocates a node using \func{malloc} during its \rdp, and before it uses the node, the thread performing the operation is neutralized.
This would cause a memory leak.


Additionally, writes to thread local data structures are not recommended.
To see why, suppose a thread maintains a thread local doubly-linked list, and also updates this list as part of the \rdp of some operation on the shared data structure.
If the thread is neutralized in middle of its update to this local list, it might corrupt the structure of the list.
\footnote{In some cases it is safe to write to thread local storage (TLS). 
For example, a thread could use TLS to maintain statistics that are \textit{supposed to} persist despite neutralization. 
Similarly, some idempotent or atomic changes to TLS should remain correct even if one is neutralized. Proceed with caution.}



\noindent \textbf{Phase 2:} reservation\todo[size=\tiny]{shall use reservation$\Phi$}. This is a \textit{conceptual stage} that does not necessarily correspond to any data structure code. However, this is where a key \nbr operation will be invoked. At this point, one must be able to identify all shared objects that will be modified by the operation in the next phase, so they can be provided to \nbr. We call these \textit{reserved} records.

\noindent \textbf{Phase 3:} \wtp ({\em write phase}). Accesses (reads/writes/CASs) to global variables, and system calls, are permitted. Accesses (including write/CAS) to shared records are permitted only if the records are \textit{reserved}. To understand what could go wrong if this restriction is violated, we need to better understand \nbr, so we will return to this restriction with an example in \secref{exwtp}.

Finally, threads not executing a data structure operation are said to be in a {\em quiescent phase} (essentially the same as phase 0).

\subsection{Overview of NBR}


In \nbr, each thread accumulates records that it has unlinked in a private buffer (or \textit{limbo bag}).
When the size of a thread $T$'s buffer exceeds a predetermined threshold, the thread sends a neutralizing signal to all other threads. 
Upon receipt of such a signal, the behaviour of a thread $T'$ depends on which phase it is executing in.

\textbf{If} $T'$ is in a quiescent phase, or preamble (Phase 0), it holds no pointers to shared records, and does not prevent $T$ from reclaiming records in its buffer.
$T'$ simply continues executing (effectively ignoring the signal).

On the other hand, \textbf{if} $T'$ is in \rdp, it may hold pointers to records in $T$'s buffer.
If $T'$ were to continue executing, it would have to prevent $T$ from reclaiming records.
Note, however, that $T'$ has not yet performed any modifications to any shared records (since it is still in \rdp).
So, $T'$ can simply discard all of its pointers (that are in its private memory), and jump back to the start of its \rdp, without leaving any shared data structures in an inconsistent state.
To implement this jump, every data structure operation invokes \func{sigsetjmp} at the start of its \rdp, which creates a \textit{checkpoint} (saving the values of all stack variables).
A thread can subsequently invoke \func{siglongjmp} to return to the last place it performed \func{sigsetjmp} (and restore the values of all stack variables).\footnote{Technically \func{sigsetjmp} saves only the current stack \textit{frame}. Stack variables defined deeper on the stack will not necessarily be saved or restored.}
It can then retry executing its \rdp, traversing a new sequence of records, starting from the \textit{root}, without any risk of accessing any records \func{free}d by $T$ (since those are no longer reachable).

The subtlety in \nbr arises when $T'$ is in \wtp.
In this case, $T'$ may hold pointers to records in the buffer of $T$.
Thus, if it continues executing, $T'$ must prevent $T$ from reclaiming these records.
Moreover, since $T'$ may have modified some shared records (but not completed its operation yet), we cannot simply restart its data structure operation, or we may leave the data structure in an inconsistent state.
So, $T'$ will \textit{not} restart its operation.
Instead, it will simply \textit{continue executing} wherever it was when it received the signal (effectively ignoring the signal).
At this point, the reader might wonder how we \textit{simultaneously avoid}: 
\begin{itemize}
    \item [\textbf{a.}] {Blocking the reclamation of $T$, and}
    \item [\textbf{b.}] {The possibility that $T'$ continues executing and it accesses a record \func{free}d 
    by $T$.}
\end{itemize}
The solution lies in the \textit{reservation} phase (Phase 2) of $T'$.
During the reservation phase of $T'$, just before it begins its \wtp, $T'$ \textit{reserves} all of the shared records it will access in its \wtp by \textit{announcing} pointers to them in a shared array.
These reservations serve a similar purpose to hazard pointers, but are quite different from HP in terms of performance and safety guarantees.
This is discussed further in \secref{usb}.
By the time $T'$ is in its \wtp (so it will ignore any neutralization signals), its reservations are visible to $T$, 
and $T$ can refer to these reservations to avoid reclaiming any of those records.

In short, operations in the \rdp discard their pointers and restart, and operations in the \wtp must have reserved them. This empowers the \textit{reclaimers} to assume that \rd{s} lose all of their pointers in response to neutralizing, and the \wt{s} lose all pointers that are not reserved. As a result, once a thread sends a neutralizing signal to all other threads, it can scan all reservations, and free any records in its buffer (limbo bag) that are not reserved.

\begin{algorithm}[t] 
    \caption{\nbr. Assumes, max number of reservations are less than the limboBag size.}\label{algo:nbr}
%
    \begin{algorithmic}[1]
    \Statex \textbf{thread local variable:}
        \State int tid \Comment{\footnotesize current thread id\normalsize}
        \State record *limboBag \label{lin:rb}\Comment{\footnotesize per-thread list of unlinked records. Maxsize:S \normalsize}
        \State bool restartable \label{lin:rst}\Comment{\footnotesize local var to track \rdp/\wtp \normalsize}
        \State record *tail \label{lin:tl}\Comment{\footnotesize Pointer to last record in limboBag \normalsize}
    \Statex
    \Statex \textbf{shared variable:}
    \State atomic<record*> reservations[N][R] \label{lin:resv} \Comment{\footnotesize N:\#threads, R:max reserved records. |R| $<<$ |S|. \normalsize }
    
    \Statex
        \Procedure{begin\rdp}{ }\label{lin:brp}
            \State \texttt{reservations[tid].clear();} \label{lin:arpc}
            \State \texttt{\Call {CAS}{\&restartable, 0, 1};}\label{lin:nz01}
        \EndProcedure

        \Procedure{end\rdp}{\{$rec_1,rec_2\cdot\cdot\cdot rec_R$\}}\label{lin:bwp}
            \State \texttt{reservations[tid] = \{$rec_1,rec_2\cdot\cdot\cdot rec_R$\};}\label{lin:arp_add}
            \State \texttt{\Call {CAS}{\&restartable, 1, 0};} \label{lin:nz10}
        \EndProcedure

        \Procedure{retire}{rec} \label{lin:func_retire}
            \If{\texttt{isLimboBagTooLarge()}} \label{lin:ioop}
                \State \texttt{\Call{signalAll}{ };}\label{lin:sa} 
                \State \texttt{\Call{reclaimFreeable}{tail};}\label{lin:rf}
            \EndIf \label{lin:ioop_eif}
            \State \texttt{limboBag[tid].append(rec);}\label{lin:re_app}
        \EndProcedure
        \Procedure{reclaimFreeable}{tail} \label{lin:func_rf}
            \State \texttt{$A$ = \Call{collectReservations}{ };}\label{lin:carp} 
            \State \texttt{$R$ = limboBag[tid].\Call{remove}{A, tail};}\label{lin:retg} 
            \State \texttt{\Call {free}{\{$R$\}};} \label{lin:free}
        \EndProcedure
    \end{algorithmic}
\end{algorithm}
\vspace{2mm}
\noindent

\subsection{Implementation of NBR}\label{sec:nbr_impl}
Algorithm 1 shows the pseudocode for \nbr.
%
Each thread collects unlinked records in its \textit{limboBag} (\lineref{rb}), and maintains
a local \nz variable that indicates whether the thread should jump back to the start of its \rdp in the event that it receives a neutralization signal (\lineref{rst}). 
We say the thread is \textit{restartable} if \nz is \textit{true} (1), and \textit{non-restartable} otherwise.
Additionally, each thread, before entering the \wtp, reserves all records it might access in a single-writer multi-reader (SWMR) \textit{reservation} array, (\lineref{resv}).
We assume the maximum number $R$ of reserved records is strictly less than maximum size $S$ of a limbo bag.

A thread in \rdp clears its reservations (if any), and then changes \nz to \emph{true} using a CAS (\Lineref{nz01}).
This CAS might initially seem strange, since it is performed on a single-writer variable and cannot fail.
The CAS prevents instruction reordering on x86-64 architectures (additional fences may be needed for more relaxed memory models).
More specifically, the goal of CAS at \lineref{nz01} is to ensure that a thread $T$ becomes restartable \textit{before} any subsequent reads of shared records.
If this CAS were simply an atomic write (rather than a read-modify-write instruction), it would be possible for $T$'s reads of shared records to be reordered before this write. In other words some reads of shared records in \rdp may appear to occur in \textit{preamble} (or previous \wtp{}) due to instruction reordering.
This could end up breaking the rule that says access to shared records is not permitted in \textit{preamble} (phase 0) as discussed in \secref{assumptions}.
As a result, the thread, which is not yet restartable, might ignore a neutralization signal and access a \func{free}d record.

Just before a thread $T$ enters a \wtp, it announces a set of reservations, and then changes \nz to \textit{false} using CAS (\Lineref{nz10}).
This CAS is used to broadcast the reservations to other threads.
More specifically, a CAS by thread $T$ at line \ref{lin:nz10} implies a memory fence, which ensures that all of the reservations (announced at the previous line~\ref{lin:arp_add}) are visible to other threads \textit{before} $T$ changes \textit{restartable} to false.
If this CAS were a simple write, it would be possible for a \rl to miss some reservations of $T$, and erroneously \func{free} those records\footnote{
Instead of using CAS, on modern x86/64 machines we can use the more efficient \func{xchg} instruction. See Section 11.5.1 of \url{https://www.amd.com/system/files/TechDocs/47414_15h_sw_opt_guide.pdf} for further details.}

In other words, the following incorrect execution may occur on x86/64 if a write is used instead of CAS: a thread $T$ reserves record $rec$ and writes 0 to restartable.
Suppose the reservations of thread $T$ remain in the processor's store buffer, and are not visible to other threads yet.
Then, another thread $T'$ sends a neutralizing signal to $T$, scans the reservations and does not see $rec$, and consequently frees $rec$.
Upon receiving the signal, $T'$ will \textit{not} restart since it has already written 0 to restartable.\footnote{
If $T$ and $T'$ are executing on different processors, then $T$ will not see the effects of any pending writes in the store buffer of $T'$, but $T'$ \textit{will} see the effects its own pending writes in order to maintain sequential consistency.}
Instead, it continues executing, and dereferences $rec$ (accessing a freed record).

The \func{retire} operation (\lineref{func_retire}) begins by checking whether the size of the limbo bag is above a predetermined threshold (32k in our experiments), at \lineref{ioop}.
If so, 
it sends a neutralizing signal to all threads using \func{signalAll} (\lineref{sa}), and then proceeds to reclaim all \textit{safe} (i.e., \textit{unreserved}) records (\lineref{rf}).
Otherwise, it simply 
adds \textit{rec} to \textit{limboBag}. 

The \func{reclaimFreeable} procedure frees all records (up to the last record pointed to by thread local pointer, \emph{tail}) in the \textit{limboBag} that are not \emph{reserved} (\lineref{func_rf}). It first scans \emph{reservations} array of all other threads and collects the reserved records in set $A$ (\lineref{carp}).
Then it removes the retired records, which are not in $A$ (set of reserved records), up to the $tail$ of the \textit{limboBag} using \func{remove(A, tail)} at \lineref{retg}. Finally, it frees the \emph{safe} set of records $R$ at \lineref{free}.


After discussing the implementation of \nbr we can now elaborate on how \rd{s}, \wt{s} and \rl{s} collaborate to achieve safe memory reclamation.

\subsubsection{Reader-reclaimer handshake}\label{sec:rrhandshake}
Each thread $T'$ at the time of \textsc{begin\rdp} saves its execution state (program counter and stack frame) using \textit{sigsetjmp} so that when it becomes restartable it can jump back to this state upon receiving a neutralizing signal. 
%
%
When a \rl $T$ sends a neutralization signal to thread $T'$, the operating system causes the control flow of $T'$ to be interrupted, so that $T'$ will immediately execute a \textit{signal handler} if $T'$ is currently running. (Otherwise, if $T'$ is not currently running, the next time it is scheduled to run it will execute the signal handler before any other steps.)
The signal handler determines whether $T'$ is restartable by reading the local \nz variable.
If the thread is restartable, then the signal handler will invoke \func{siglongjmp} and jump back to the start of the \rdp (so it is as if $T'$ never started the \rdp{}). 

This behaviour represents a sort of two-step \textit{handshake} between \textit{readers} (threads in \rdp) and \textit{reclaimers} (threads executing lines 16 and 17 in \func{retire}) to avoid scenarios where a reader might access a \func{free}d record. 
A \rl guarantees that before \emph{reclaiming} any of its \emph{unlinked} records it will signal all threads, and all \rd{s} guarantee that they will relinquish any reference to unsafe records when they receive a neutralization signal.

\subsubsection{Writers handshake} \label{sec:whandshake}
(1) Each \rl signals all threads before starting to reclaim any records.
When a \wt receives a signal, it executes a \emph{signalHandler} that determines the thread is non-restartable, and immediately returns. 
The \rl then goes on to reclaim its \emph{limboBag} (\lineref{rf}), except for any reserved records contained therein, independently from the actions of the \wt.

This is safe because a \wt, before entering into the \wtp, \textit{reserves} all of the shared records it might access in its \wtp 
(\lineref{arp_add}).
Thus, (2) the \wt guarantees to the \rl that, although it will not restart its data structure operation, it will only access \emph{reserved} records.
The (3) \rl, in turn, guarantees it will scan all announcements after signaling and before reclaiming the contents of its \emph{limboBag}, and will consequently avoid reclaiming any records that will be accessed by the \wt in its \wtp.

This three-step handshake formed by (1), (2) and (3) avoids scenarios where a \wt might access a \func{free}d record. 
Crucially, all \wt{s} atomically ensure that their reserved records are visible to the \rl at the moment they become non-restartable.
In turn, \rl{s} scan reservations \textit{after} sending neutralization signals (at which point any thread that does not restart has already made its reservations visible). 


\subsection{Revisiting the \wtp restriction}
\label{sec:exwtp}
In this section, we will trace an incorrect execution that could occur if a thread accesses any record that is not reserved \textit{before} entering the \wtp.

Suppose a thread $T$ is in a \wtp, and sleeps just before it accesses a shared record $rec$, which it has \textit{not} reserved.
Then, another thread $T'$ sends a neutralization signal to $T$ using \func{signalAll}.
Next, $T'$ scans the \emph{reservations} array of the thread $T$.
$T$ did not reserve $rec$ so $T'$ will not find $rec$ in $T$'s reserved records (which violates the writers handshake, \secref{whandshake}).
Therefore, $T'$ will assume that $rec$ can be freed safely, and will do so.
Finally, $T$ wakes up and proceeds with its \textit{unsafe} access of $rec$.

\myparagraph{Ensuring reclamation can occur.}
The total number of records that can be reserved over all threads 
must be strictly smaller than the limboBag capacity, in order to ensure that threads \textit{can} reclaim records whenever the limboBag is full.
In practice, most data structures require few reservations.
For example in our experiments, operations in the lazylist~\cite{heller2005lazy} required at most 2 reservations, and the Harris list~\cite{harris2001pragmatic}, DGT~\cite{david2015asynchronized}, and (a,b)-trees ~\cite{brown2017techniques} at most 3.


\section{NBR+}
\label{sec:nbrp}
Next, we explain a performance issue with \nbr which motivated us to design an improved version called \nbrp.
%
Signals on linux trigger page-fault routines and a switch from user to kernel mode that incurs significant overhead. 
Therefore, it is desirable to send as few signals as possible (while maintaining high reclamation throughput). 
However, every time a thread reclaims records from its \textit{limboBag}, NBR requires the thread to send signals to \textit{all} other threads.
This induces a \textit{relaxed grace period} (RGP): A time interval [t, t'] during which each thread is neutralized due to a reclamation event triggered by some reclaimer thread.
In NBR, every thread induces a RGP every time it tries to reclaim its \textit{limboBag}.
As a result, in order for all $n$ threads to reclaim their \textit{limboBag}s, $n(n-1)$ signals must be sent.
The need to send $O(n^2)$ signals to allow all $n$ threads to reclaim memory can severely limit performance. 
Naturally, we would like to improve this. 

Suppose, in NBR, at some time $t$, a thread sends $n-1$ signals to other threads so it can reclaim its \textit{limboBag}.
This causes all of the other $n-1$ threads to discard any unreserved references to shared records.
Meaning, at time $t$, the (unreserved) records in the limbo bags of \textbf{all threads} are safe to free.
Therefore, if somehow we could propagate this information that a RGP has occurred due to some thread $T$, then all other threads could piggyback on $T$ to partially or completely reclaim their own limbo bags without sending signals of their own.
In other words, in the best case, all $n$ participating threads could reclaim memory after detecting exactly one RGP, induced by sending a total of $n-1$ signals. 


\myparagraph{Overview of NBR+}
The key insight in \nbrp is that when a reclaimer sends neutralization signals to \textit{all} threads, \textit{all} threads discard their pointers to unreserved records, and thus \textit{all} threads can potentially reclaim some records in their \textit{limboBag}s.
%
This suggests a design wherein each thread (1) passively detects a RGP by observing signals sent by another thread, and (2) determines which records in its \textit{limboBag} were unlinked \textit{before} the RGP (i.e., are safe to reclaim).

\subsection{Implementation of NBR+}

\label{sec:nbrp_impl}
We explain the design of \nbrp by building our exposition around three main design challenges. 
\begin{enumerate}
    \item [(C1)] {When should a thread start tracking other threads' signals to detect a \rgp?}
    \item [(C2)] {How can a thread \textit{recognize} that a \rgp has occurred?}
    \item [(C3)] {Once a thread recognises that a \rgp has occurred how should it determine which records in its \emph{limboBag} are safe to reclaim?}
\end{enumerate}

\begin{algorithm}[t]
\small
\caption{\nbrp: Only variables that differ from \nbr are shown here. \nbrp includes all variables and procedures of \algoref{nbr}. The \func{retire} operation is different in \nbrp.}\label{algo:nbrp} 


    \begin{algorithmic}[1]
    \Statex \textbf{thread local variable:}
        \State int scanTS[N]; \Comment{N = \#threads. P = set of processes}
        \State bool firstLoWmEntryFlag = true;
        \State record* bookmarkTail; 
    \Statex
    \Statex \textbf{shared variable:}
    \State atomic<int> announceTS[N];
    \Statex    
        \Procedure{retire}{rec}
            \If{isAtHiWm()} \label{lin:iahw}
                \State \texttt{\Call{FAA}{\&announceTS[tid],1};}\Comment{\small \rgp begin \normalsize} \label{lin:afaa1}
                \State \texttt{\Call{signalAll()}{}} \label{lin:sigall}
                \State \texttt{\Call {FAA}{\&announceTS[tid],1};}\Comment{\small \rgp end \normalsize} \label{lin:afaa2}                
                \State \texttt{\Call{reclaimFreeable}{tail};} \label{lin:recfr}
                \State \texttt{\Call{cleanUp()}{};}\label{lin:iahwend}
            \ElsIf{isAtLoWm()} \label{lin:ialw}
                \If{firstLoWmEntryFlag}
                    \State \texttt{bookmarkedTail $=$ tail;} \label{lin:bmtail}
                    \State \texttt{scanTS[tid] $=$ \Call {$scanAnnounceTS()$}{}} \label{lin:readscants}
                \EndIf
                \For{each otid $\in$ P}\Comment{\small otid: other thread's id in P. \normalsize}\label{lin:attemptfreebeg}
                    \If{\small announceTS[otid]$\geq$scanTS[tid][otid]+2 \normalsize} \label{lin:pback}
                        \State \texttt{\Call{reclaimFreeable}{bookmarkTail};} \label{lin:prf}
                        \State \texttt{\Call{cleanUp()}{};} \label{lin:cluplw}
                        \State \texttt{break;}
                    \EndIf            
                \EndFor \label{lin:attemptfreeend}
            \EndIf \label{lin:ialw_end}
            \State \texttt{limboBag[tid].append(rec);} \label{lin:apprec}
        \EndProcedure
        
        \Procedure{cleanUp}{} \label{lin:clup}
            \State \texttt{firstLoWmEntryFlag $=$ 1;}
        \EndProcedure
    \end{algorithmic}
\end{algorithm}


As a solution to \textbf{(C1)}, each thread in \nbrp, in addition to watching the \lb size to determine when it becomes \textit{too large} (triggering neutralization), also determines when the \lb size crosses a predetermined threshold called the \lw (e.g., one half full or one quarter full). 
If a thread's \lb is full, we say that thread is at the \hw. 
If a thread's \lb keeps growing without reclamation it will first cross the \lw and then hit the \hw. 
As shown in \algoref{nbrp}, a thread determines whether it has passed the \hw or \lw using procedures \func{isAtHiWm} (\lineref{iahw}) and \func{isAtLoWm} (\lineref{ialw}). 
Once a thread has passed the \lw, it begins recording and analyzing information about signals sent by other threads to detect RGPs.

To tackle \textbf{(C2)}, a \rl at the \lw (who wants to detect a \rgp) must perform a sort of handshake with another \rl at the \hw (who triggers a \rgp).
\nbrp implements this handshake using per-thread single-writer multi-reader timestamps (similar to vector clocks).

Whenever a \rl{} hits the \hw, it first increments its timestamp (to an odd value) to indicate that it is \textit{currently broadcasting signals} (\lineref{afaa1}).
This denotes the \textit{beginning} of a \rgp.
It then sends signals to all threads, and increments its timestamp again (to an even value) to indicate that it has \textit{finished broadcasting signals} (\lineref{afaa2}).
This denotes the \textit{end} of the \rgp. 

Whenever a \rl $T$ passes the \lw, it 
collects and saves the current timestamps of all threads (\lineref{readscants}), as well as the current \textit{tail} pointer of its \textit{limboBag} (\lineref{bmtail}), so it can remember precisely which records it had unlinked \textit{before} it reached its \lw.
$T$ then periodically collects the timestamps of all threads, comparing the new values it sees to the original values it saw when it passed the \lw (\lineref{attemptfreebeg} - \lineref{attemptfreeend}).
(To obtain high performance, we amortize the overhead of scanning \textit{announceTS} over many \func{retire} operations.)
It continues to do this until it either detects a RGP or hits the \hw itself (and sends signals to induce its own RGP).
Observe that, after $T$ hits its \lw, if the timestamp of any thread changes from one even number to another even number, then that thread has both \textit{begun and finished} sending signals to \textit{all} threads since $T$ hit the \lw.
Thus, $T$ can identify that a \rgp has occurred since it hit its \lw, solving (C2). 

Finally, to tackle \textbf{(C3)}, observe that T saves the last record ($tail$ of its limboBag) it had retired before entering the \lw at \lineref{bmtail}.
If T successfully observes a \rgp as explained in the solution to (C2), then all threads would either have discarded or reserved all their private references to the records in $T$'s limboBag up to the saved $bookmarkTail$. 
Thus, $T$ can invoke \func{reclaimFreeable} to free all unreserved records up to the $bookmarkTail$ (\lineref{prf}). solving (C3).

\func{cleanUp()} (\lineref{clup}) method is used to set \emph{firstLoWmEntryFlag} after a thread reclaims either at \lw (\lineref{cluplw}) or at \hw (\lineref{iahwend}). 

A thread that has not reached the \lw or the \hw simply continues to append any retired records to its \textit{limboBag} (\lineref{apprec}).

\ignore{
A reclaimer that has reached the \hw reclaims exactly as it would in \nbr, meaning it broadcasts a neutralization signal to all threads (\lineref{sigall}), causing an \rgp, and then 
reclaims safe records (skipping any reserved records, (\lineref{recfr})).

Execution on the \lw path (\lineref{ialw}--\ref{lin:ialw_end}) monitors the signals that have been sent in the system to try to avoid sending signals in the future, incurring some overhead that is amortized over multiple data structure operations.

To solve \textbf{(C2)}, a thread that passes the \lw has to periodically check for a \emph{relaxed grace period}.
In order to be able to reclaim memory without sending any signals, the reclaimer at the \lw must perform a 
handshake with another reclaimer at the \hw (who \textit{is} sending signals). 
}

At first it may appear that a thread $T$ can reclaim its \textit{limboBag} as soon as it receives a neutralizing signal from a \rl thread $T'$.
However, the receipt of a single signal is not enough for $T$ to safely reclaim memory.
To safely reclaim the set $R$ of records in its \lb up to its $bookmarkTail$, $T$ needs to know that \textit{all} threads have been neutralized \textit{since $T$ retired the records in $R$}.
Otherwise, some other thread may still have a pointer to a record in $R$.

Let us discuss an example of what can go wrong if a thread reclaims its \textit{limboBag} after it receives a single signal.
Consider a system with three threads $T1$, $T2$ and $T3$.
Suppose $T1$ is at its \hw, $T2$ is at its \lw and $T3$ holds a private reference to a record $rec$ that is in $T2$'s \textit{limboBag}.
$T1$, being at its \hw, begins neutralizing all threads one by one.
First, it sends neutralizing signal to $T2$ (starting a \rgp).
$T2$, upon receiving the signal, reclaims its \textit{limboBag} including $rec$.
Note, that $T1$ hasn't neutralized $T3$ yet, meaning a \rgp has not yet occurred.
Now, if $T3$ accesses $rec$, a \textit{use-after-free} error would occur.
To prevent this, $T2$ should not reclaim the contents of its \textit{limboBag} unless $T1$ \textit{completes} the \rgp by neutralizing $T3$ (preventing $T3$ from doing this unsafe access). 
The crucial point is that $T2$ must detect the \textit{start} and  \textit{end} of a RGP to know that it can safely reclaim records in its \lb.

\subsection{Applicability}
\label{sec:appl}

\nbr(+)\footnote{We will simply write \nbr in this section with the understanding that the entire discussion applies identically to \nbrp.} 
naturally applies to many concurrent data structures that have synchronization-free searches followed by update(s) because in such data structures searches and updates correspond to the \rdp and the \wtp of \nbr, respectively (as shown in \figref{3phase}).
Thus, to apply \nbr one just needs to invoke \textsc{begin\rdp} before the start of the search and \textsc{end\rdp} before starting the update(s).
For example, in the lazy-list of Heller et al.~\cite{heller2005lazy}, the \rdp of an operation would begin with the start of the search for target records and the \wtp would consist of the locking and validation of target records followed by any modifications to them.


Certain other lock-free data structures exhibit a pattern where searches, in an operation, perform \textit{auxiliary update(s)} followed by intended update(s).
Such an operation has a \textit{sequence} of \emph{read-write phases}. 
For example, in Harris's lock-free list~\cite{harris2001pragmatic}, while searching the list towards a target location, a thread may modify the list by unlinking any \textit{marked} (logically deleted) records it encounters. 
Then, once it arrives at the target location, it performs 
the operation's intended modification. 

Since \nbr is designed for a single \rdp and \wtp, applying it carelessly to such a data structure 
could break the requirement that, after entering a \wtp, no new records are discovered.
(This would be unsafe, because it breaks the \wt{s} handshake.)
For instance, in such a data structure, if we enter a \wtp to perform an \textit{auxiliary update}, \nbr would be stuck in the \wtp, unable to obtain new pointers (that have not yet been reserved) to continue its traversal. 

That said, \nbr \textit{can} be applied in some data structures that would require \textit{multiple read/write} phases, provided that each consecutive pair of read and write phases obey the requirements set out in Section~\ref{sec:assumptions}.

\myparagraph{Example: Harris list.}
\algoref{knbr_eg} shows how \nbr can be used with the Harris list~\cite{harris2001pragmatic}, \textit{despite} the fact that this list has auxiliary updates.
We hope the reader can follow our exposition on the Harris list, and infer how \nbr could be applied to more sophisticated data structures with similar design patterns (such as Brown's ABTree \cite{brown2017techniques}, which appears in our experiments).


\begin{algorithm}[t]
\small
    \caption{Integration of \nbr with Harris list\cite{harris2001pragmatic} with multiple read/write phases $(\rdp \space\space \wtp)^+$.}\label{algo:knbr_eg}
\begin{lstlisting}[]
bool insert(key) {
  Node *right_node, *left_node;
  do{
      right_node = search (key, &left_node);
      if((right_node!=tail) && (right_node.key==key)) 
        return false; 
      Node *new_node = |\textcolor{red}{new Node(key)}|;
      new_node.next = right_node; 
      if (|\textcolor{red}{CAS}|(&(left_node.next), right_node, new_node))
        return true;
  }while (true)
}

Node* search(key, Node** left_node) {|\label{lin:ksrch}|
  Node *left_node_next, *right_node;
  search_again:
  do {
      |\textcolor{red}{begin\rdp{}();}|    |\label{lin:ksrchrdp}|    
      Node *t = head;
      Node *t_next = head.next;
      do{
          if(!is_marked_reference(t_next)){
            (*left_node) = t;
            left_node_next = t_next;
          }
          t = get_unmarked_reference(t_next);
          if (t == tail) break;
          t_next = t.next;
      }while(is_marked_reference(t_next) or (t.key<search_key));
      right_node = t;
      |\textcolor{red}{end\rdp{}(left\_node, right\_node);}|   |\label{lin:ksrchwtp}| 
    
      if (left_node_next == right_node) 
        if ((right_node != tail) && is_marked_reference(right_node.next))
          goto search_again;
        else 
          return right_node; |\label{lin:ksrchret}| 
      if (|\textcolor{red}{CAS}|(&(left_node.next), left_node_next, right_node)) 
        if ((right_node != tail) && is_marked_reference(right_node.next))
          goto search_again;
        else 
          return right_node;
  } while(true);
}
\end{lstlisting}
\label{fig:alg3}
\end{algorithm}

To understand how \nbr behaves when applied to the Harris list, suppose the initial list configuration is $L$: $1_f \Longrightarrow 2_f \Longrightarrow 3_t \Longrightarrow 4_f \Longrightarrow 6_f \Longrightarrow 10_f$, where each node is represented as \emph{key$_{marked}$} (where $marked$ is [$t$]rue or [$f$]alse).
Now, suppose a thread $T$ performs \emph{Ins:insert(9)}, starting with an invocation of \func{search()}.
This invocation of \func{search()} starts a \rdp (\lineref{ksrchrdp}) and begins traversing $L$.
Starting from $\langle pred, curr \rangle = \langle1_f,2_f\rangle$ the thread observes $\langle pred, curr \rangle = \langle2_f,3_t\rangle$, where $curr=3_t$ is marked.
To remove marked node $3_t$ (an auxiliary \textit{helping} update), $T$ enters a \wtp (\lineref{ksrchwtp}) and changes the next pointer of $2_f$ to $4_f$, yielding the list configuration: $1_f \Longrightarrow 2_f \Longrightarrow 4_f \Longrightarrow 6_f \Longrightarrow 10_f$.
Moving forward, $T$'s \func{search()} will enter a \textit{second} \rdp (\lineref{ksrchrdp}), and traverse the list again, \textit{starting from the root}. 
As $T$ now obtain pointers to \textit{new nodes} (which would be impossible with only a single \rdp and \wtp), we must argue that it doesn't access any \func{free}d nodes.
However, this is straightforward, since it is again traversing \textit{from the root} discarding any references from previous \textit{read-write} phases.
(From the perspective of SMR, it is as if $T$ has simply started a new data structure operation.)

Now, suppose $T$ is \textit{neutralized} by a concurrent \rl while it is in this second \rdp.
Upon receipt of a neutralization signal, $T$ will jump back to the beginning of its \textit{second} \rdp, and restart its search, once again, from the \textit{root}.
Note that neutralizing does not affect the lock-free progress guarantee, since a thread sends neutralization signals only after performing many successful deletion operations.
Suppose $T$ eventually performs a \rdp that reaches the nodes $\langle pred, curr \rangle = \langle6_f,10_f\rangle$ where it should perform its modification.
$T$ will then enter a final \wtp and insert $9_f$ after returning (\lineref{ksrchret}) from the \func{search()}, yielding $L$: $1_f \Longrightarrow 2_f \Longrightarrow 4_f \Longrightarrow 6_f \Longrightarrow 9_f \Longrightarrow 10_f$.

\myparagraph{Limitation: restarting from the root.}
In order for \nbr to be safe, it is \textit{crucial that Ins forgets all pointers and restarts from the \textit{root} every time it begins a new \rdp}.
Intuitively, this is because each new read phase is effectively a new data structure operation---all pointers are forgotten when the new \rdp begins.
If it attempts to continue searching from somewhere in the middle of the list, perhaps by resuming its search from a shared node $R$ that was reserved by the previous \wtp, then \textit{Ins} could easily dereference a \func{free}d node.
To see why, note that, although $R$ cannot be \func{free}d (since it is reserved), the nodes that it points to are \textit{not} necessarily reserved, and so they could be \func{free}d.
Thus, as soon as \textit{Ins} follows any pointer starting from $R$, it could access a \func{free}d node and crash.

\begin{table*}[t]
\centering
\begin{tabular}{lllllll}
\textbf{Source}                                                         & \textbf{Data structure}              & \textbf{Sync. type} & \textbf{NBR+}      & \textbf{EBR} & \textbf{DEBRA+} & \textbf{HP/TS/IBR/HE/WFE/HY/QSense} \\
LL05\cite{heller2005lazy}             & linked list                 & opt. locks      & Yes       & Yes & No     & No (similar to \cite{brown2015reclaiming})                \\
\rowcolor[rgb]{0.753,0.753,0.753} HL01\cite{harris2001pragmatic}                                          & linked list                 & lock-free       & Yes       & Yes & *      & Yes                \\
HM04\cite{michael2004hazard}          & linked list                 & lock-free       & No        & Yes & *      & Yes                \\

\rowcolor[rgb]{0.753,0.753,0.753} \rowcolor[rgb]{0.753,0.753,0.753}DVY14a\cite{drachsler2014practical}    & partially external BST      & locks           & **        & Yes & No     & No \cite{brown2015reclaiming}                \\
EFRB10\cite{ellen2010non}             & external BST                & lock-free       & Yes       & Yes & *      & No \cite{brown2015reclaiming}                \\
\rowcolor[rgb]{0.753,0.753,0.753} NM14\cite{natarajan2014fast}                                            & external BST                & lock-free       & Yes       & Yes & *      & No \cite{brown2015reclaiming}                 \\
EFRB14\cite{ellen2014amortized}                                         & external BST                & lock-free       & No        & Yes & *      & No \cite{brown2015reclaiming}                \\
\rowcolor[rgb]{0.753,0.753,0.753} DGT15\cite{david2015asynchronized}                                      & external BST                & ticket locks    & Yes       & Yes & No     & No (no marks, cannot validate HP)                \\
HJ12\cite{howley2012non}                                                & internal BST                & lock-free       & Yes       & Yes & *      & No (similar to \cite{brown2015reclaiming})                \\
\rowcolor[rgb]{0.753,0.753,0.753} RM15\cite{ramachandran2015fast}       & internal BST                & lock-free       & No        & Yes & No     & No (similar to \cite{brown2015reclaiming})                \\

BCCO10\cite{bronson2010practical}     & partially external AVL      & opt. locks      & No        & Yes & No     & Yes                \\
\rowcolor[rgb]{0.753,0.753,0.753} DVY14b\cite{drachsler2014practical}                                     & partially external AVL      & locks           & No        & Yes & No     & No \cite{brown2015reclaiming}                \\
HL17\cite{he2017deletion}             & external relaxed AVL tree   & lock-free       & Yes       & Yes & Yes    & No (similar to \cite{brown2015reclaiming})                 \\
\rowcolor[rgb]{0.753,0.753,0.753} B17b\cite{brown2017techniques}                                          & external AVL                & lock-free       & Yes       & Yes & Yes    & No \cite{brown2015reclaiming}                \\

S13\cite{shafiei2013non}              & patricia trie               & lock-free       & Yes       & Yes & *      & No \cite{brown2015reclaiming}                \\
\rowcolor[rgb]{0.753,0.753,0.753}BER14\cite{brown2014general}                                                                   & external chromatic tree     & lock-free       & Yes       & Yes & Yes    & No \cite{brown2015reclaiming}                \\
B17a\cite{brown2017techniques}        & external (a,b)-tree         & lock-free       & Yes       & Yes & Yes    & No \cite{brown2015reclaiming}                \\
\rowcolor[rgb]{0.753,0.753,0.753}BPA20\cite{brown2020non}                                                & external interpolation tree & lock-free       & No        & Yes & No     & No (similar to \cite{brown2015reclaiming})                \\
\end{tabular}
\caption{\textbf{Applicability of SMR algorithms.}
A detailed explanation of the contents of this table can be found in \secref{apnappl} of appendix.
*It appears likely that DEBRA+ is compatible, but one must design non-trivial data structure specific recovery code. 
**This is likely possible if code is restructured to reserve all relevant nodes before acquiring any locks.
}
\label{tab:appltab}
\vspace{-6mm}
\end{table*}

\myparagraph{Compatible data structures.}
There are numerous concurrent data structures in the literature with multiple \textit{read-write phases} that \textit{do} restart from the \textit{root} after any \textit{auxiliary updates}, and hence are natural candidates for pairing with \nbr.
For example, Harris' list~\cite{harris2001pragmatic}, Brown's lock-free ABTree, chromatic tree and AVL tree (B17) \cite{brown2017techniques}, the lock-free binary search tree of Natarajan et~al.~\cite{natarajan2014fast},
and many more \cite{ellen2010non,howley2012non,shafiei2013non,he2017deletion}.
Among these, we used the harris list and the ABTree in our experiments. The Harris list appears in \figref{apnhlist} of \secref{apnexp}.

\myparagraph{Semi-compatible data structures.}
The need to restart from the root at the start of each \rdp suggests that \nbr cannot be used with the data structures like the Harris-Michael list \cite{michael2004hazard}, and some search trees \cite{ellen2014amortized, drachsler2014practical, brown2020non, ramachandran2015castle,bronson2010practical}, wherein the searches (\rdp) after each \textit{auxiliary update} (\wtp) do not start from the \textit{root}.
However, we could potentially use \nbr with such data structures if we were to modify the operations so they restart \textit{from the root} after any auxiliary updates.
Depending on the data structure, this might break the progress guarantee (for example changing a wait-free algorithm into a lock-free one, or necessitating a new amortized complexity analysis~\cite{ellen2014amortized}), or it might simply add overhead. 

For some data structures, the overhead of restarting from the \textit{root} may be quite low in practice, and forcing operations to restart from the root may be a reasonable solution.
(The cost of restarting from the root is studied in our experiments.)
For example, in Harris-Michael list, in high contention scenarios where $k$ threads all contend on an auxiliary CAS to unlink the same marked node, all threads except for the one that succeeds this CAS would already restart from the root~\cite{michael2004hazard} anyway!
If we modify this list so threads always restart from the root, in this high contention scenario, $k$ threads must restart instead of $k-1$.
Incidentally, by doing this, we essentially obtain the Harris list \cite{harris2001pragmatic}, in which all threads contending on the auxiliary CAS already restart from the \textit{root}.
(In low contention scenarios the way we restart should not affect performance significantly.)

Furthermore, in search trees, assuming a uniform distribution of accesses, threads tend to spread out in the tree, so average contention is quite low.
This suggests that, when a thread encounters contention, the performance difference between restarting from the root and continuing a traversal from an ancestor will be small in many workloads.

\myparagraph{Incompatible data structures.}
We are aware of a few data structures that are incompatible with (or would require extensive code changes to work with) \nbr.
Two concurrent implementations of a relaxed-balance AVL tree appear in~\cite{bronson2010practical,drachsler2014practical}.
In each of these implementations, after a key is inserted, rotations must be performed to rebalance the tree. These rotations are performed starting at the bottom of the tree, possibly continuing all the way to the root (traversing upwards using parent pointers).
In the process of performing these rotations, a thread may encounter many new nodes that were \textit{not} traversed as part of the initial search in the insert operation.
In order to use \nbr with these algorithms, one would need to rewrite the implementations to perform a new search from the root \textit{for each rotation}. 

A recent lock-free interpolation search tree~\cite{brown2020non} also appears to be incompatible.
For example, in this algorithm, entire subtrees are periodically rebuilt to maintain balance, and during this process, threads \textit{mark} all nodes in the subtree, one by one, alternating between steps that \textit{mark} a node and \textit{discover} a new node (without restarting from the root in between).
It is not clear how one could transform this algorithm into the form required by \nbr.
(Note, however, that neither DEBRA+ nor HPs can be used with this data structure, either. We are not aware of \textit{any} SMR algorithm with bounded garbage that is compatible with this tree.)

\myparagraph{Comparing with other SMR algorithms.}
\nbr can be used with many data structures that other SMR algorithms like DEBRA+ and HP (and variants of HPs, including HE, IBR, WFE, ThreadScan, HY and QSense) are incompatible with~\cite{heller2005lazy, ellen2010non, natarajan2014fast, david2015asynchronized, howley2012non}.
There are also some data structures that are compatible with other SMR algorithms but not \nbr~\cite{michael2004hazard,bronson2010practical}. See \tabref{appltab} for an overview. 
A detailed analysis of the table's contents, and an exposition of how \nbr can be applied to these data structures, is relegated to \secref{apnappl} in the appendix accompanying this paper.


\subsection{Ease of use}
\label{sec:usb}
\begin{figure*}[ht]
\noindent
\begin{minipage}{.33\textwidth}
\begin{lstlisting}[basicstyle=\tiny\ttfamily,frame=tlrb,morekeywords={*, recl_start_op,recl_end_op, recl_retire,...},numbers=none, frame=L]{Name}
void OP_DEBRA()
{
|\colorbox{light-gray}{recl\_start\_op}|
RETRY:
    pred=head; curr=pred.next;
    while (key |$\leq$| curr.key) {
        pred=curr;
        curr=cur.next;
    }
    
    if (key == curr.key) {
        return false;
    }
    
    lock(pred);
    lock(curr);
    if (!validate()) {
        unlock(pred); unlock(curr);
        goto RETRY;
    }
    
    do update
    unlock(pred); unlock(curr);
|\colorbox{light-gray}{recl\_end\_op \hspace{2mm}}|
}
\end{lstlisting}
\subcaption{DEBRA}
\label{fig:usbdebra}
\end{minipage}\hfill
\begin{minipage}{.33\textwidth}
\begin{lstlisting}[basicstyle=\tiny\ttfamily,frame=tlrb,morekeywords={*, start_op,end_op, recl_retire, upgrade_to_write_phase, save_for_write_phase,...},numbers=none, frame=L]{Name}|\label{fig:usbnbrp}|
void OP_NBR+()
{
RETRY:
|\colorbox{light-gray}{begin\rdp}|
    pred = head; curr = pred.next;
    while(key |$\leq$| curr.key) {
        pred=curr;
        curr=cur.next;
    }

|\colorbox{light-gray}{end\rdp}|
    if (key == curr.key) {
        return false;
    }
    
    lock(pred); lock(curr);
    if (!validate()) {
        unlock(pred); unlock(curr);
        goto RETRY;
    }
    
    do update
    unlock(pred); unlock(curr);
}
\end{lstlisting}
\subcaption{\nbrp}
\label{fig:usbnbr}
\end{minipage}\hfill
\begin{minipage}{.33\textwidth}
\begin{lstlisting}[basicstyle=\tiny\ttfamily,frame=tlrb,morekeywords={*, recl_start_op,recl_end_op, recl_retire, upgrade_to_write_phase, save_for_write_phase,...},numbers=none, frame=L]{Name}
void OP_HP()
{
RETRY:
    pred = head; curr = pred.next;
    |\colorbox{light-gray}{protect(curr) RETRY on fail}|;
    while (key |$\leq$| curr.key) {
        |\colorbox{light-gray}{unprotect(pred)}|;
        pred=curr; 
        curr=cur.next;
        |\colorbox{light-gray}{protect(curr) RETRY on fail}|;
    }
    if (key == curr.key) {
        |\colorbox{light-gray}{unprotect(pred)}; \colorbox{light-gray}{unprotect(curr)}|;
        return false;
    }
    lock(pred); lock(curr);
    if (!validate()) {
        |\colorbox{light-gray}{unprotect(pred)}; \colorbox{light-gray}{unprotect(curr)}|;
        unlock(pred);unlock(curr);
        goto RETRY;
    }
    do update
    |\colorbox{light-gray}{unprotect(pred)}; \colorbox{light-gray}{unprotect(curr)}|;
    unlock(pred);unlock(curr);
}
\end{lstlisting}
\vspace{-4mm}
\subcaption{HP}
\label{fig:usbhp}
\end{minipage}
\vspace*{-4mm}
\caption{Complexity of using \debra , \nbr and \hp on a lazy list. \debra $<<$ \nbr $<<$ \hp.}
\label{fig:usb}
\vspace*{-3mm}
\end{figure*}
\figref{usb} compares the difficulty of using \hp, \nbr and \debra in the insert operation of the lazy list of Heller et~al.~\cite{heller2005lazy}.
As \figref{usbhp} demonstrates, \hp is cumbersome to use because it requires a programmer to protect every record by \textit{announcing} hazard pointers, using a store/load fence or $xchg$ instruction to ensure that each announcement is visible in a timely manner by other threads, validating that the announced record is still safe before dereferencing it, and restarting if validation fails.
Programmers also need to unprotect records that they will no longer dereference, further increasing the need for intrusive code changes. 

On the contrary, applying \nbr to a data structure operation is, intuitively, similar to performing two-phased locking, in the sense that the primary difficulty revolves around identifying where the \wtp should begin, and which records it will access.
The programmer just needs to invoke \func{begin\rdp} before the operation accesses its first shared record, in this example, at the start of the traversal for target records. 
Then s/he must invoke \func{end\rdp} before modifying any shared records.
In this example, the \wtp begins just before the lock acquisition on pred.
If there are no modifications to be performed in an operation, for example, in the \func{contains} operation of the lazy-list, then the programmer can simply invoke \func{end\rdp} before returning from the operation.

\debra is simplest as it requires programmers to invoke just two functions corresponding to the start and the end of a data structure operation 
(\figref{usbdebra}). 

In terms of programmer effort, \nbr finds a middleground between DEBRA and HPs. 
Although \nbr is slightly more involved than DEBRA, we believe that the benefits due to \nbr's bounded garbage property and better performance outweigh the extra effort of identifying which shared records will be modified by the \wtp and where in the code to invoke \func{end\rdp}. 

Just to give readers a quantitative view of the amount of programming effort needed to use HP and NBR we measured number of extra reclamation related lines of code needed to be written in our implementation of \func{insert(), delete() and contains()} methods for the lazylist and DGT. We observed that \nbr required only 10 extra lines of code in comparison to 30 extra lines of code needed to use HP.

As mentioned earlier, in \nbr, before a thread enters a \wtp, it must reserve all the records that will be accessed in the \wtp.
In some data structures it might not be possible to determine \textit{precisely} which records \textit{will} be accessed in the \wtp.
For example, in a tree, an operation may decide \textit{during} the write phase whether to modify the left or right child pointer. 
To apply \nbr in such a tree, one can simply reserve \textit{both} pointers. 
(Nevertheless there may be some data structures where it is infeasible to reserve \textit{all} of the records that might be accessed in a \wtp.)

\section{Correctness}
\label{sec:corr}
We show that \nbr and \nbrp are both safe and have bounded garbage.
Interested readers can find detailed proofs in \secref{apncor} of the appendix accompanying this paper.
\begin{lemma}[NBR is Safe]
\label{lem:nbr}
Reclaimer threads in \nbr only reclaim records that are safe (i.e., that no other thread has access to). 
\end{lemma}
Intuitively, this follows from the handshakes between \rd{s}, \wt{s} and \rl{s}.




\begin{lemma}[\nbrp is safe]
Reclaimer threads in \nbrp only reclaim records that are safe. 
\end{lemma}
Intuitively, in addition to the handshakes between \rd{s}, \wt{s} and \rl{s} the fact that \rl{s} at \lw only reclaim a record iff they observe a \rgp ensures that \nbrp is safe.  



\begin{lemma}[Both \nbr and \nbrp have bounded garbage] The number of unlinked, unreclaimed records is bounded. 
\end{lemma}
This follows from the fact that all \rl{s} forcefully reclaims their \emph{limboBag} when it is full except the reserved records which are bounded. See appendix for formal proof.  



\section{Experimental Evaluation}
\label{sec:eval}
\myparagraph{Setup:} We used a quad-socket Intel Xeon Platinum 8160 machine running at 2.1GHz with 192 hardware threads and 384~GiB memory having shared L3 cache (33.79~MiB) on Ubuntu 18.04 with GCC/G++ version 7.4.0.

All algorithms used in the experiments were implemented in the Setbench~\cite{brown2020non} benchmark compiled with \texttt{-O3} optimization, and used \emph{jemalloc} as the memory allocator~\cite{evans2006scalable}. We perform four kinds of experiments:
\begin{enumerate}
    \item [(E1):] {Studies throughput over different thread counts and workloads to understand \nbr(+)'s performance and scalability [P1].}
    \item [(E2):] {Studies peak memory usage of \nbrp to show the advantage of bounded garbage [P2].}
    \item [(E3):] {Studies the impact of contention on performance [P4].}
    \item [(E4):] {Studies the impact of modifying a data structure to restart from the root to make it work with \nbrp.} 
\end{enumerate}
\vspace{-2mm}

\begin{figure*}[htbp]
     \begin{minipage}{\textwidth}
        \begin{subfigure}{\textwidth}
            \includegraphics[width=0.33\linewidth, height=4cm, keepaspectratio]{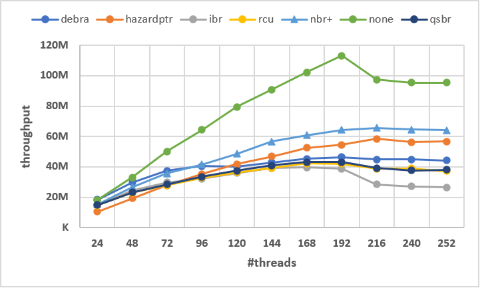}\hfill
            \includegraphics[width=0.33\linewidth, height=4cm, keepaspectratio]{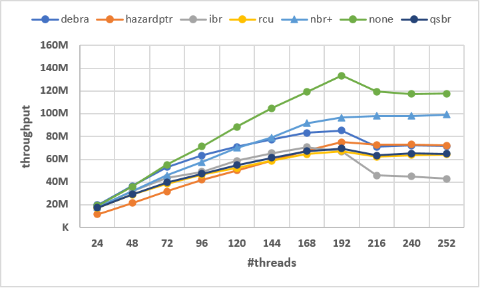}\hfill
            \includegraphics[width=0.33\linewidth, height=4cm, keepaspectratio]{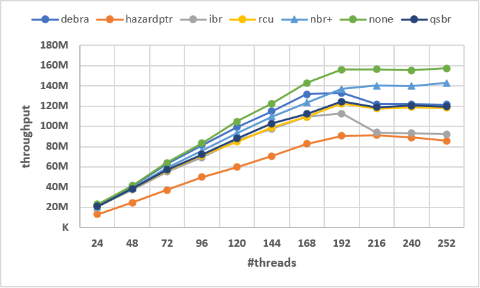}\hfill
            \caption{DGT-tree. Left: 50i-50d. Middle: 25i-25d. Right: 5i-5d. Key range size:2M.}
            \label{fig:tree}
        \end{subfigure}
        \begin{subfigure}{\textwidth}
            \includegraphics[width=0.33\linewidth, height=4cm, keepaspectratio]{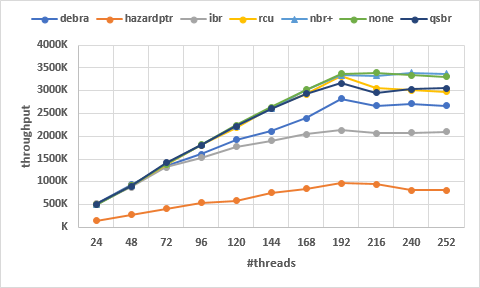}\hfill
            \includegraphics[width=0.33\linewidth, height=4cm, keepaspectratio]{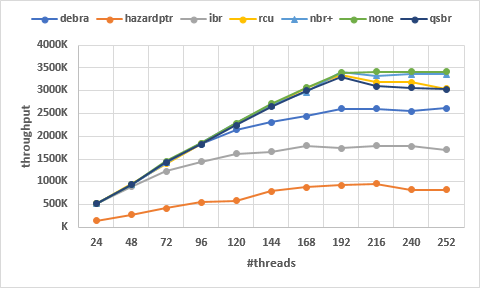}\hfill
            \includegraphics[width=0.33\linewidth, height=4cm, keepaspectratio]{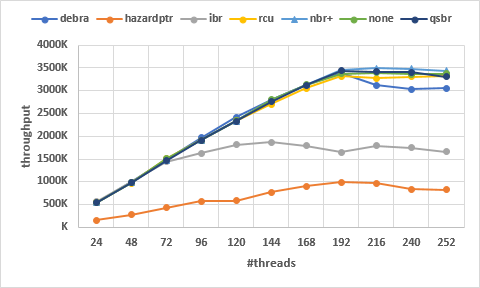}\hfill
            \caption{Lazy linked-list. Left: 50i-50d. Middle: 25i-25d. Right: 5i-5d. Key range size:20K.}
            \label{fig:list}
        \end{subfigure}
     \end{minipage}
\vspace*{-3mm}
    \caption{E1: Evaluation of throughput. Y axis: throughput in million operations per second. X axis: \#threads. 
    }
    \label{fig:exp}
\end{figure*}

\begin{figure*}[htbp]
\begin{minipage}{.66\textwidth}
        \begin{subfigure}{\textwidth}
            \includegraphics[width=0.5\linewidth, height=5cm, keepaspectratio]{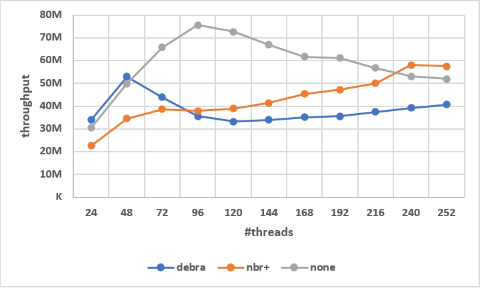}\hfill
            \includegraphics[width=0.5\linewidth, height=5cm, keepaspectratio]{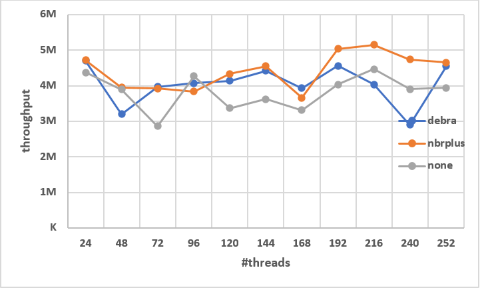}\hfill
            \caption{E3: Brown's ABTree. Left: 50i-50d. Key range size: 2M. Right: 50i-50d. Key range size:200. Comparison with Harris list is in appendix.}
            \label{fig:abt}
        \end{subfigure}
        \begin{subfigure}{\textwidth}
            \includegraphics[width=0.5\linewidth, height=5cm, keepaspectratio]{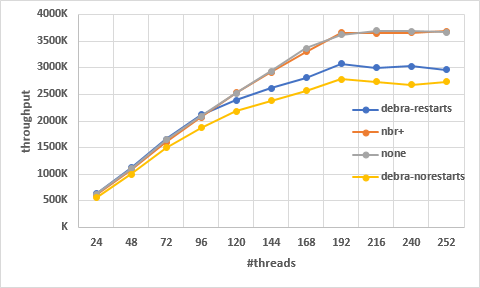}\hfill
            \includegraphics[width=0.5\linewidth, height=5cm, keepaspectratio]{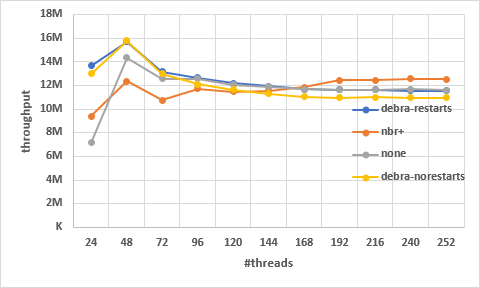}\hfill
            \caption{E4: Lock-free Harris-Michael list. Left: 50i-50d. Key range size:20K. Right: 50i-50d. Key range size:200. The debra-norestarts use HMlist\cite{michael2004hazard} whereas \nbrp, debra-restarts and none use modified HMList that restarts from the root.}
            \label{fig:hl}
        \end{subfigure}
\end{minipage}%
\hfill
\begin{minipage}{.33\textwidth}
    \begin{subfigure}{\linewidth}
        \includegraphics[width=6.1cm, height=6.5cm, keepaspectratio]{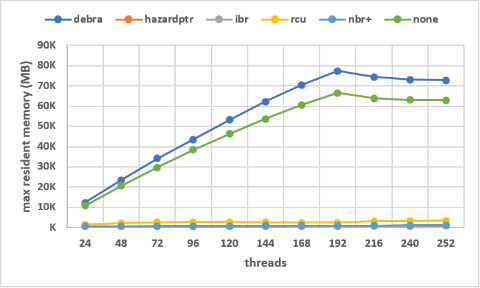}\hfill
        \caption{E2: With stalled threads.}
        \label{fig:figure1}
    \end{subfigure}
    \vfill
    \begin{subfigure}{\textwidth}
        \includegraphics[width=6.1cm, height=6.5cm, keepaspectratio]{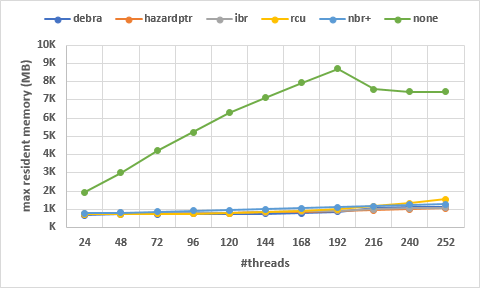}\hfill
        \caption{E2: NO stalled threads.}
        \label{fig:figure2}
    \end{subfigure}
\end{minipage}%
\vspace*{-3mm}
\caption{Fig. \textbf{(a), (b)}: Evaluation of throughput for data structures with multiple read-write phases. Y axis: millions of ops/sec. Fig. \textbf{(c), (d)}: Y axis: Max Resident Memory (MB), Workload: 50i-50d. DGT-tree. Key range size: 2M.} 
\label{fig:exp_knbr}
\vspace{-2mm}
\end{figure*}

For (E1) we picked the lazy-list~\cite{heller2005lazy} and DGT~\cite{david2015asynchronized} as representative of lists and trees,
to evaluate \nbrp against QSBR, RCU, DEBRA, and the 2geibr variant of interval based reclamation (IBR)~\cite{wen2018interval}, hazard pointers (HP) and a leaky implementation (none).
(We adapted QSBR, RCU and 2geibr (IBR) from the IBR benchmark, 
integrating them into Setbench to ensure a fair comparison.)
For (E2) we compared peak memory usage for each of the aforementioned reclamation algorithms using DGT. 
For (E3) we compared \nbr with DEBRA and the leaky implementation (none) using the ABTree data structure \cite[chapter 8]{brown2017techniques} for very large and small data structure size.
Finally, for (E4), we modified the Harris-Michael list (HMList) such that every \rdp restarts from the root. 
This allowed us to use \nbrp to reclaim memory for this list.
To understand the impact of these restarts, we also used DEBRA to reclaim memory for this modified HMList (labeled \textit{DEBRA-restarts} in our graphs), as well as the original HMList (DEBRA-norestarts).

Reported results are obtained by averaging data over 3 timed trials, each lasting 5 seconds, for each thread count in \{24, 48, 72, 96, 120, 144, 168, 192, 216, 240, 252\}, and each data structure.
We used a key range size of 2~M and 20~K for trees and lists, respectively.
Each execution starts by prefilling the data structure to half of the key range size, i.e., 1~M for trees and 10~K for lists.



For each of (E1), (E2), and (E3) we subject \nbrp to exhaustive evaluation by running it on three workload profiles, (1) \textit{update-intensive} 50\% inserts and 50\% deletes, (2) \textit{balanced} 25\% inserts, 25\% deletes and 50\% searches, and (3) \textit{search-intensive} 90\% searches, 5\% inserts and 5\% deletes.
(We also run with oversubscription---i.e., more threads than logical processors, to establish P4.)


\myparagraph{Discussion}
(E1) results for the DGT tree and lazy linked list appear in \Figref{exp}.
\nbrp matches or outperforms its competitors for nearly all data points.
In the tree, it surpasses the next best algorithm, DEBRA, by up to $38\%$ and $12\%$  (\Figref{tree}, update-intensive and balanced workloads, resp.) and is comparable to DEBRA in search-intensive workloads where it outperforms the next best algorithm by up to 10\% (\Figref{tree} search-intensive workload). 



In these graphs, DEBRA performs better than NBR+ for low thread counts, but NBR+ outperforms it after 96 threads in update intensive workloads (\Figref{tree}, leftmost plot), and after 120 threads in the balanced workload (\Figref{tree}, center).
The two algorithms are comparable in the search-intensive workload (\Figref{tree}, rightmost plot).
The poor performance of DEBRA at higher thread counts could be attributed to the infrequent advancement of epochs by slow threads, which leads to halting of regular reclamation of limbo bags. We call this the \textit{delayed thread vulnerability}.
As a result, the limbo bags of all threads keep on growing until the slow thread finally catches up and announces the required epoch.

The delayed thread vulnerability leads to the accumulation of a large number of retired records waiting to be reclaimed. 
Once the current epoch is announced by the slow thread, all threads reclaim their large limbo bags, which leads to a \textit{reclamation burst}. 
This harms the overall throughput, as reclamation bursts can bottleneck the underlying allocator (jemalloc in our experiments) by increasing contention and triggering slow/fallback code paths.
The probability of threads getting delayed increases with high thread counts and update-intensive workloads. 

Furthermore, one may notice that the thread count where \nbrp overtakes DEBRA is different in the \textit{update intensive} and \textit{search-intensive} workloads.
This could be attributed to the fact that the overhead of burst reclamation sets in at lower thread counts for \textit{update-intensive} workloads than in workloads with infrequent updates.

HP outperforms the other EBR variants in update-intensive workloads (\Figref{tree}, leftmost plot) but they appear to be slow in the search-intensive workload (\Figref{tree}, rightmost plot). This can be attributed to the fact that overhead due to the delayed thread vulnerability dominates the overhead of per-record fencing in HP for update-intensive workloads.
Whereas, for search-intensive workloads the overhead due to delayed threads is lesser than that of per-record fences in HPs.
Also, as one can observe in \Figref{list}, in the lazylist \nbrp is comparable to \rcu, \qsbr, and \debra, and performs better than \hp (by 2x) and \ibr (by more than $50\%$) across all workloads when oversubscribed [P1, P4].\footnote{The poor performance of HPs is partly due to the cost of \textit{mfence}, which could be replaced with the more efficient \func{xchg} (see Section 11.5.1 in \url{https://www.amd.com/system/files/TechDocs/47414_15h_sw_opt_guide.pdf}).
We tried this optimization in separate experiments, and while it did increase throughput, HPs remained significantly slower than \nbrp.}


In E2 (\Figref{figure1} and ~\ref{fig:figure2}), we validate the bounded garbage property of \nbrp [P2] by measuring the peak memory usage of all reclamation algorithms when a thread is stalled (\Figref{figure1}) and when no thread is stalled (\Figref{figure2}).
Each trial is run for 25 seconds.
During the entire length of the experiment (25 seconds) one thread is made to begin a data structure operation and then sleep. 
As expected, since \debra and \rcu do not have bounded garbage, they exhibit an increase in peak memory usage in presence of a stalled thread (\Figref{figure1}) while \nbrp, \hp, and \ibr variants maintain approximately the same peak memory usage. 

In E3 (\Figref{abt}) we evaluate throughput of \nbr with a data structure in which operations already restart from the root whenever a new read phase would be started, namely the Brown ABTree\cite[chapter 8]{brown2017techniques}. 
We design our experiments to explore two disparate usage scenarios.
First, we want to study reasonably large data structure (key range size 2~M) wherein we hypothesize restarts would be inexpensive due to low contention (leftmost in \Figref{exp_knbr}).
Second, we want to study extremely small data structures (key range size 200) where restarting from the head node will occur as frequently as possible due to high contention (rightmost in \Figref{exp_knbr}). 


In the ABTree, \nbrp is faster than the other SMR algorithms in the low contention scenario (especially at high thread counts).
Moreover, in the high contention scenario, \nbrp is comparable to DEBRA, suggesting that \nbrp introduces relatively little overhead in practice due to restarting from the root in \rdp (\Figref{abt}). For more detailed comparison of ABTree across different data structure sizes and workload types refer to \Figref{apnabtree} in appendix. We also extend E3 with Harris list in \Figref{apnhlist} of appendix.

In E4 (\Figref{hl}), for low contention, \nbrp (with forced restarts from the root) is faster than both DEBRA-restarts and DEBRA-norestarts.
Shockingly, DEBRA-restarts is actually slightly \textit{faster} than DEBRA-norestarts.
The only code difference between these two implementations is an extra restart from the root in one case of DEBRA-restarts.
It appears that forced restarts actually have a backoff-like (contention managing) effect.
We found that adding restarts lowered L3 cache misses slightly, which is what we would expect from contention management.
Under high contention, \nbrp is comparable to DEBRA-restarts and DEBRA-norestarts.
This suggests that the cost of added restarts should be reasonable in practice. Additional experiments evaluating the behaviour of \nbr across different data structures sizes appear in appendix.



\noindent
\textbf{Code:}
The latest version of our source code can be found on gitlab (\url{https://gitlab.com/aajayssingh/nbr_setbench}). 

\section{Conclusions}
\label{sec:conclusion}
In this paper, we presented \nbr, a \smr algorithm that is a hybrid between EBR and a limited form of HPBR, and which uses POSIX signals to bound unreclaimed garbage.
\nbr is simpler to use than the most similar hybrid algorithm, DEBRA+, while supporting a large class of data structures, some of which are not supported by DEBRA+ (nor other popular SMR algorithms).
We also developed an optimized version of \nbr called \nbrp that achieves similar throughput with fewer signals by passively observing signals being sent in the system to optimistically detect relaxed grace periods.
Our experiments demonstrate that \nbrp meets or exceeds the performance of the state of the art in SMR algorithms in typical benchmark conditions, while minimizing performance degradation in oversubscribed workloads.


\begin{acks}
This work was supported by the Natural Sciences and Engineering Research Council of Canada (NSERC) under the grants: CRDPJ 539431-19, DGECR-2019-00048, RGPIN-2019-04227 and RGPIN-04512-2018. The John R. Evans Leaders Fund and the Ontario Research Fund (CFI):38512, Waterloo Huawei Joint Innovation Lab project ``Scalable Infrastructure for Next Generation Data Management Systems'', and the University of Waterloo. We would also like to thank the reviewers for their insightful comments. 
\end{acks}

\bibliographystyle{ACM-Reference-Format}
\bibliography{references}


\begin{thebibliography}{46}


\ifx \showCODEN    \undefined \def \showCODEN     #1{\unskip}     \fi
\ifx \showDOI      \undefined \def \showDOI       #1{#1}\fi
\ifx \showISBNx    \undefined \def \showISBNx     #1{\unskip}     \fi
\ifx \showISBNxiii \undefined \def \showISBNxiii  #1{\unskip}     \fi
\ifx \showISSN     \undefined \def \showISSN      #1{\unskip}     \fi
\ifx \showLCCN     \undefined \def \showLCCN      #1{\unskip}     \fi
\ifx \shownote     \undefined \def \shownote      #1{#1}          \fi
\ifx \showarticletitle \undefined \def \showarticletitle #1{#1}   \fi
\ifx \showURL      \undefined \def \showURL       {\relax}        \fi
\providecommand\bibfield[2]{#2}
\providecommand\bibinfo[2]{#2}
\providecommand\natexlab[1]{#1}
\providecommand\showeprint[2][]{arXiv:#2}

\bibitem[\protect\citeauthoryear{Afek, Kaplan, Korenfeld, Morrison, and
  Tarjan}{Afek et~al\mbox{.}}{2014}]%
        {afek2014cb}
\bibfield{author}{\bibinfo{person}{Yehuda Afek}, \bibinfo{person}{Haim Kaplan},
  \bibinfo{person}{Boris Korenfeld}, \bibinfo{person}{Adam Morrison}, {and}
  \bibinfo{person}{Robert~E Tarjan}.} \bibinfo{year}{2014}\natexlab{}.
\newblock \showarticletitle{The CB tree: a practical concurrent self-adjusting
  search tree}.
\newblock \bibinfo{journal}{\emph{Distributed computing}} \bibinfo{volume}{27},
  \bibinfo{number}{6} (\bibinfo{year}{2014}), \bibinfo{pages}{393--417}.
\newblock


\bibitem[\protect\citeauthoryear{Alistarh, Eugster, Herlihy, Matveev, and
  Shavit}{Alistarh et~al\mbox{.}}{2014}]%
        {alistarh2014stacktrack}
\bibfield{author}{\bibinfo{person}{Dan Alistarh}, \bibinfo{person}{Patrick
  Eugster}, \bibinfo{person}{Maurice Herlihy}, \bibinfo{person}{Alexander
  Matveev}, {and} \bibinfo{person}{Nir Shavit}.}
  \bibinfo{year}{2014}\natexlab{}.
\newblock \showarticletitle{Stacktrack: An automated transactional approach to
  concurrent memory reclamation}. In \bibinfo{booktitle}{\emph{Proceedings of
  the Ninth European Conference on Computer Systems}}. \bibinfo{pages}{1--14}.
\newblock


\bibitem[\protect\citeauthoryear{Alistarh, Leiserson, Matveev, and
  Shavit}{Alistarh et~al\mbox{.}}{2017}]%
        {alistarh2017forkscan}
\bibfield{author}{\bibinfo{person}{Dan Alistarh}, \bibinfo{person}{William
  Leiserson}, \bibinfo{person}{Alexander Matveev}, {and} \bibinfo{person}{Nir
  Shavit}.} \bibinfo{year}{2017}\natexlab{}.
\newblock \showarticletitle{Forkscan: Conservative memory reclamation for
  modern operating systems}. In \bibinfo{booktitle}{\emph{Proceedings of the
  Twelfth European Conference on Computer Systems}}. \bibinfo{pages}{483--498}.
\newblock


\bibitem[\protect\citeauthoryear{Alistarh, Leiserson, Matveev, and
  Shavit}{Alistarh et~al\mbox{.}}{2018}]%
        {alistarh2018threadscan}
\bibfield{author}{\bibinfo{person}{Dan Alistarh}, \bibinfo{person}{William
  Leiserson}, \bibinfo{person}{Alexander Matveev}, {and} \bibinfo{person}{Nir
  Shavit}.} \bibinfo{year}{2018}\natexlab{}.
\newblock \showarticletitle{Threadscan: Automatic and scalable memory
  reclamation}.
\newblock \bibinfo{journal}{\emph{ACM Transactions on Parallel Computing
  (TOPC)}} \bibinfo{volume}{4}, \bibinfo{number}{4} (\bibinfo{year}{2018}),
  \bibinfo{pages}{1--18}.
\newblock


\bibitem[\protect\citeauthoryear{Arbel-Raviv, Brown, and Morrison}{Arbel-Raviv
  et~al\mbox{.}}{2018}]%
        {arbel2018getting}
\bibfield{author}{\bibinfo{person}{Maya Arbel-Raviv}, \bibinfo{person}{Trevor
  Brown}, {and} \bibinfo{person}{Adam Morrison}.}
  \bibinfo{year}{2018}\natexlab{}.
\newblock \showarticletitle{Getting to the Root of Concurrent Binary Search
  Tree Performance}. In \bibinfo{booktitle}{\emph{Proceedings of the 2018
  USENIX Conference on Usenix Annual Technical Conference}} (Boston, MA, USA)
  \emph{(\bibinfo{series}{USENIX ATC '18})}. \bibinfo{publisher}{USENIX
  Association}, \bibinfo{address}{USA}, \bibinfo{pages}{295–306}.
\newblock
\showISBNx{9781931971447}


\bibitem[\protect\citeauthoryear{Balmau, Guerraoui, Herlihy, and
  Zablotchi}{Balmau et~al\mbox{.}}{2016}]%
        {balmau2016fast}
\bibfield{author}{\bibinfo{person}{Oana Balmau}, \bibinfo{person}{Rachid
  Guerraoui}, \bibinfo{person}{Maurice Herlihy}, {and} \bibinfo{person}{Igor
  Zablotchi}.} \bibinfo{year}{2016}\natexlab{}.
\newblock \showarticletitle{Fast and robust memory reclamation for concurrent
  data structures}. In \bibinfo{booktitle}{\emph{Proceedings of the 28th ACM
  Symposium on Parallelism in Algorithms and Architectures}}.
  \bibinfo{pages}{349--359}.
\newblock


\bibitem[\protect\citeauthoryear{Blelloch and Wei}{Blelloch and Wei}{2020}]%
        {blelloch2020concurrent}
\bibfield{author}{\bibinfo{person}{Guy~E Blelloch} {and}
  \bibinfo{person}{Yuanhao Wei}.} \bibinfo{year}{2020}\natexlab{}.
\newblock \showarticletitle{Concurrent Reference Counting and Resource
  Management in Wait-free Constant Time}.
\newblock \bibinfo{journal}{\emph{arXiv preprint arXiv:2002.07053}}
  (\bibinfo{year}{2020}).
\newblock


\bibitem[\protect\citeauthoryear{Braginsky, Kogan, and Petrank}{Braginsky
  et~al\mbox{.}}{2013}]%
        {braginsky2013drop}
\bibfield{author}{\bibinfo{person}{Anastasia Braginsky}, \bibinfo{person}{Alex
  Kogan}, {and} \bibinfo{person}{Erez Petrank}.}
  \bibinfo{year}{2013}\natexlab{}.
\newblock \showarticletitle{Drop the anchor: lightweight memory management for
  non-blocking data structures}. In \bibinfo{booktitle}{\emph{Proceedings of
  the twenty-fifth annual ACM symposium on Parallelism in algorithms and
  architectures}}. \bibinfo{pages}{33--42}.
\newblock


\bibitem[\protect\citeauthoryear{Bronson, Casper, Chafi, and Olukotun}{Bronson
  et~al\mbox{.}}{2010}]%
        {bronson2010practical}
\bibfield{author}{\bibinfo{person}{Nathan~G Bronson}, \bibinfo{person}{Jared
  Casper}, \bibinfo{person}{Hassan Chafi}, {and} \bibinfo{person}{Kunle
  Olukotun}.} \bibinfo{year}{2010}\natexlab{}.
\newblock \showarticletitle{A practical concurrent binary search tree}.
\newblock \bibinfo{journal}{\emph{ACM Sigplan Notices}} \bibinfo{volume}{45},
  \bibinfo{number}{5} (\bibinfo{year}{2010}), \bibinfo{pages}{257--268}.
\newblock


\bibitem[\protect\citeauthoryear{Brown}{Brown}{2017}]%
        {brown2017techniques}
\bibfield{author}{\bibinfo{person}{Trevor Brown}.}
  \bibinfo{year}{2017}\natexlab{}.
\newblock \showarticletitle{Techniques for Constructing Efficient Lock-free
  Data Structures}.
\newblock \bibinfo{journal}{\emph{arXiv preprint arXiv:1712.05406}}
  (\bibinfo{year}{2017}).
\newblock


\bibitem[\protect\citeauthoryear{Brown, Ellen, and Ruppert}{Brown
  et~al\mbox{.}}{2014a}]%
        {Brown:2014}
\bibfield{author}{\bibinfo{person}{Trevor Brown}, \bibinfo{person}{Faith
  Ellen}, {and} \bibinfo{person}{Eric Ruppert}.}
  \bibinfo{year}{2014}\natexlab{a}.
\newblock \showarticletitle{A General Technique for Non-blocking Trees}. In
  \bibinfo{booktitle}{\emph{Proceedings of the 19th ACM SIGPLAN Symposium on
  Principles and Practice of Parallel Programming}}
  \emph{(\bibinfo{series}{PPoPP '14})}. \bibinfo{pages}{329--342}.
\newblock
\newblock
\shownote{Full version available from \url{http://tbrown.pro}.}


\bibitem[\protect\citeauthoryear{Brown, Ellen, and Ruppert}{Brown
  et~al\mbox{.}}{2014b}]%
        {brown2014general}
\bibfield{author}{\bibinfo{person}{Trevor Brown}, \bibinfo{person}{Faith
  Ellen}, {and} \bibinfo{person}{Eric Ruppert}.}
  \bibinfo{year}{2014}\natexlab{b}.
\newblock \showarticletitle{A general technique for non-blocking trees}. In
  \bibinfo{booktitle}{\emph{Proceedings of the 19th ACM SIGPLAN symposium on
  Principles and practice of parallel programming}}. \bibinfo{pages}{329--342}.
\newblock


\bibitem[\protect\citeauthoryear{Brown, Prokopec, and Alistarh}{Brown
  et~al\mbox{.}}{2020}]%
        {brown2020non}
\bibfield{author}{\bibinfo{person}{Trevor Brown}, \bibinfo{person}{Aleksandar
  Prokopec}, {and} \bibinfo{person}{Dan Alistarh}.}
  \bibinfo{year}{2020}\natexlab{}.
\newblock \showarticletitle{Non-blocking interpolation search trees with
  doubly-logarithmic running time}. In \bibinfo{booktitle}{\emph{Proceedings of
  the 25th ACM SIGPLAN Symposium on Principles and Practice of Parallel
  Programming}}. \bibinfo{pages}{276--291}.
\newblock


\bibitem[\protect\citeauthoryear{Brown}{Brown}{2015}]%
        {brown2015reclaiming}
\bibfield{author}{\bibinfo{person}{Trevor~Alexander Brown}.}
  \bibinfo{year}{2015}\natexlab{}.
\newblock \showarticletitle{Reclaiming memory for lock-free data structures:
  There has to be a better way}. In \bibinfo{booktitle}{\emph{Proceedings of
  the 2015 ACM Symposium on Principles of Distributed Computing}}.
  \bibinfo{pages}{261--270}.
\newblock


\bibitem[\protect\citeauthoryear{Cohen}{Cohen}{2018}]%
        {cohen2018every}
\bibfield{author}{\bibinfo{person}{Nachshon Cohen}.}
  \bibinfo{year}{2018}\natexlab{}.
\newblock \showarticletitle{Every data structure deserves lock-free memory
  reclamation}.
\newblock \bibinfo{journal}{\emph{Proceedings of the ACM on Programming
  Languages}} \bibinfo{volume}{2}, \bibinfo{number}{OOPSLA}
  (\bibinfo{year}{2018}), \bibinfo{pages}{1--24}.
\newblock


\bibitem[\protect\citeauthoryear{Cohen and Petrank}{Cohen and Petrank}{2015a}]%
        {cohen2015automatic}
\bibfield{author}{\bibinfo{person}{Nachshon Cohen} {and} \bibinfo{person}{Erez
  Petrank}.} \bibinfo{year}{2015}\natexlab{a}.
\newblock \showarticletitle{Automatic memory reclamation for lock-free data
  structures}.
\newblock \bibinfo{journal}{\emph{ACM SIGPLAN Notices}} \bibinfo{volume}{50},
  \bibinfo{number}{10} (\bibinfo{year}{2015}), \bibinfo{pages}{260--279}.
\newblock


\bibitem[\protect\citeauthoryear{Cohen and Petrank}{Cohen and Petrank}{2015b}]%
        {cohen2015efficient}
\bibfield{author}{\bibinfo{person}{Nachshon Cohen} {and} \bibinfo{person}{Erez
  Petrank}.} \bibinfo{year}{2015}\natexlab{b}.
\newblock \showarticletitle{Efficient memory management for lock-free data
  structures with optimistic access}. In \bibinfo{booktitle}{\emph{Proceedings
  of the 27th ACM symposium on Parallelism in Algorithms and Architectures}}.
  \bibinfo{pages}{254--263}.
\newblock


\bibitem[\protect\citeauthoryear{David, Guerraoui, and Trigonakis}{David
  et~al\mbox{.}}{2015}]%
        {david2015asynchronized}
\bibfield{author}{\bibinfo{person}{Tudor David}, \bibinfo{person}{Rachid
  Guerraoui}, {and} \bibinfo{person}{Vasileios Trigonakis}.}
  \bibinfo{year}{2015}\natexlab{}.
\newblock \showarticletitle{Asynchronized concurrency: The secret to scaling
  concurrent search data structures}.
\newblock \bibinfo{journal}{\emph{ACM SIGARCH Computer Architecture News}}
  \bibinfo{volume}{43}, \bibinfo{number}{1} (\bibinfo{year}{2015}),
  \bibinfo{pages}{631--644}.
\newblock


\bibitem[\protect\citeauthoryear{Detlefs, Martin, Moir, and Steele~Jr}{Detlefs
  et~al\mbox{.}}{2002}]%
        {detlefs2002lock}
\bibfield{author}{\bibinfo{person}{David~L Detlefs}, \bibinfo{person}{Paul~A
  Martin}, \bibinfo{person}{Mark Moir}, {and} \bibinfo{person}{Guy~L
  Steele~Jr}.} \bibinfo{year}{2002}\natexlab{}.
\newblock \showarticletitle{Lock-free reference counting}.
\newblock \bibinfo{journal}{\emph{Distributed Computing}} \bibinfo{volume}{15},
  \bibinfo{number}{4} (\bibinfo{year}{2002}), \bibinfo{pages}{255--271}.
\newblock


\bibitem[\protect\citeauthoryear{Dice, Herlihy, and Kogan}{Dice
  et~al\mbox{.}}{2016}]%
        {dice2016fast}
\bibfield{author}{\bibinfo{person}{Dave Dice}, \bibinfo{person}{Maurice
  Herlihy}, {and} \bibinfo{person}{Alex Kogan}.}
  \bibinfo{year}{2016}\natexlab{}.
\newblock \showarticletitle{Fast non-intrusive memory reclamation for
  highly-concurrent data structures}. In \bibinfo{booktitle}{\emph{Proceedings
  of the 2016 ACM SIGPLAN International Symposium on Memory Management}}.
  \bibinfo{pages}{36--45}.
\newblock


\bibitem[\protect\citeauthoryear{Drachsler, Vechev, and Yahav}{Drachsler
  et~al\mbox{.}}{2014}]%
        {drachsler2014practical}
\bibfield{author}{\bibinfo{person}{Dana Drachsler}, \bibinfo{person}{Martin
  Vechev}, {and} \bibinfo{person}{Eran Yahav}.}
  \bibinfo{year}{2014}\natexlab{}.
\newblock \showarticletitle{Practical concurrent binary search trees via
  logical ordering}. In \bibinfo{booktitle}{\emph{Proceedings of the 19th ACM
  SIGPLAN symposium on Principles and practice of parallel programming}}.
  \bibinfo{pages}{343--356}.
\newblock


\bibitem[\protect\citeauthoryear{Dragojevi{\'c}, Herlihy, Lev, and
  Moir}{Dragojevi{\'c} et~al\mbox{.}}{2011}]%
        {dragojevic2011power}
\bibfield{author}{\bibinfo{person}{Aleksandar Dragojevi{\'c}},
  \bibinfo{person}{Maurice Herlihy}, \bibinfo{person}{Yossi Lev}, {and}
  \bibinfo{person}{Mark Moir}.} \bibinfo{year}{2011}\natexlab{}.
\newblock \showarticletitle{On the power of hardware transactional memory to
  simplify memory management}. In \bibinfo{booktitle}{\emph{Proceedings of the
  30th annual ACM SIGACT-SIGOPS symposium on Principles of distributed
  computing}}. \bibinfo{pages}{99--108}.
\newblock


\bibitem[\protect\citeauthoryear{Ellen, Fatourou, Helga, and Ruppert}{Ellen
  et~al\mbox{.}}{2014}]%
        {ellen2014amortized}
\bibfield{author}{\bibinfo{person}{Faith Ellen}, \bibinfo{person}{Panagiota
  Fatourou}, \bibinfo{person}{Joanna Helga}, {and} \bibinfo{person}{Eric
  Ruppert}.} \bibinfo{year}{2014}\natexlab{}.
\newblock \showarticletitle{The amortized complexity of non-blocking binary
  search trees}. In \bibinfo{booktitle}{\emph{Proceedings of the 2014 ACM
  symposium on Principles of distributed computing}}.
  \bibinfo{pages}{332--340}.
\newblock


\bibitem[\protect\citeauthoryear{Ellen, Fatourou, Ruppert, and van
  Breugel}{Ellen et~al\mbox{.}}{2010}]%
        {ellen2010non}
\bibfield{author}{\bibinfo{person}{Faith Ellen}, \bibinfo{person}{Panagiota
  Fatourou}, \bibinfo{person}{Eric Ruppert}, {and} \bibinfo{person}{Franck van
  Breugel}.} \bibinfo{year}{2010}\natexlab{}.
\newblock \showarticletitle{Non-blocking binary search trees}. In
  \bibinfo{booktitle}{\emph{Proceedings of the 29th ACM SIGACT-SIGOPS symposium
  on Principles of distributed computing}}. \bibinfo{pages}{131--140}.
\newblock


\bibitem[\protect\citeauthoryear{Evans}{Evans}{2006}]%
        {evans2006scalable}
\bibfield{author}{\bibinfo{person}{Jason Evans}.}
  \bibinfo{year}{2006}\natexlab{}.
\newblock \showarticletitle{A scalable concurrent malloc (3) implementation for
  FreeBSD}. In \bibinfo{booktitle}{\emph{Proc. of the bsdcan conference,
  ottawa, canada}}.
\newblock


\bibitem[\protect\citeauthoryear{Fatourou, Papavasileiou, and Ruppert}{Fatourou
  et~al\mbox{.}}{2019}]%
        {fatourou2019persistent}
\bibfield{author}{\bibinfo{person}{Panagiota Fatourou}, \bibinfo{person}{Elias
  Papavasileiou}, {and} \bibinfo{person}{Eric Ruppert}.}
  \bibinfo{year}{2019}\natexlab{}.
\newblock \showarticletitle{Persistent non-blocking binary search trees
  supporting wait-free range queries}. In \bibinfo{booktitle}{\emph{The 31st
  ACM Symposium on Parallelism in Algorithms and Architectures}}.
  \bibinfo{pages}{275--286}.
\newblock


\bibitem[\protect\citeauthoryear{Fraser}{Fraser}{2004}]%
        {fraser2004practical}
\bibfield{author}{\bibinfo{person}{Keir Fraser}.}
  \bibinfo{year}{2004}\natexlab{}.
\newblock \bibinfo{booktitle}{\emph{Practical lock-freedom}}.
\newblock \bibinfo{type}{{T}echnical {R}eport}.
  \bibinfo{institution}{University of Cambridge, Computer Laboratory}.
\newblock


\bibitem[\protect\citeauthoryear{Gidenstam, Papatriantafilou, Sundell, and
  Tsigas}{Gidenstam et~al\mbox{.}}{2008}]%
        {gidenstam2008efficient}
\bibfield{author}{\bibinfo{person}{Anders Gidenstam}, \bibinfo{person}{Marina
  Papatriantafilou}, \bibinfo{person}{H{\aa}kan Sundell}, {and}
  \bibinfo{person}{Philippas Tsigas}.} \bibinfo{year}{2008}\natexlab{}.
\newblock \showarticletitle{Efficient and reliable lock-free memory reclamation
  based on reference counting}.
\newblock \bibinfo{journal}{\emph{IEEE Transactions on Parallel and Distributed
  Systems}} \bibinfo{volume}{20}, \bibinfo{number}{8} (\bibinfo{year}{2008}),
  \bibinfo{pages}{1173--1187}.
\newblock


\bibitem[\protect\citeauthoryear{Harris}{Harris}{2001}]%
        {harris2001pragmatic}
\bibfield{author}{\bibinfo{person}{Timothy~L Harris}.}
  \bibinfo{year}{2001}\natexlab{}.
\newblock \showarticletitle{A pragmatic implementation of non-blocking
  linked-lists}. In \bibinfo{booktitle}{\emph{International Symposium on
  Distributed Computing}}. Springer, \bibinfo{pages}{300--314}.
\newblock


\bibitem[\protect\citeauthoryear{Hart, McKenney, Brown, and Walpole}{Hart
  et~al\mbox{.}}{2007}]%
        {hart2007performance}
\bibfield{author}{\bibinfo{person}{Thomas~E Hart}, \bibinfo{person}{Paul~E
  McKenney}, \bibinfo{person}{Angela~Demke Brown}, {and}
  \bibinfo{person}{Jonathan Walpole}.} \bibinfo{year}{2007}\natexlab{}.
\newblock \showarticletitle{Performance of memory reclamation for lockless
  synchronization}.
\newblock \bibinfo{journal}{\emph{J. Parallel and Distrib. Comput.}}
  \bibinfo{volume}{67}, \bibinfo{number}{12} (\bibinfo{year}{2007}),
  \bibinfo{pages}{1270--1285}.
\newblock


\bibitem[\protect\citeauthoryear{He and Li}{He and Li}{2017}]%
        {he2017deletion}
\bibfield{author}{\bibinfo{person}{Meng He} {and} \bibinfo{person}{Mengdu Li}.}
  \bibinfo{year}{2017}\natexlab{}.
\newblock \showarticletitle{Deletion without rebalancing in non-blocking binary
  search trees}. In \bibinfo{booktitle}{\emph{20th International Conference on
  Principles of Distributed Systems (OPODIS 2016)}}. Schloss
  Dagstuhl-Leibniz-Zentrum fuer Informatik.
\newblock


\bibitem[\protect\citeauthoryear{Heller, Herlihy, Luchangco, Moir, Scherer, and
  Shavit}{Heller et~al\mbox{.}}{2005}]%
        {heller2005lazy}
\bibfield{author}{\bibinfo{person}{Steve Heller}, \bibinfo{person}{Maurice
  Herlihy}, \bibinfo{person}{Victor Luchangco}, \bibinfo{person}{Mark Moir},
  \bibinfo{person}{William~N Scherer}, {and} \bibinfo{person}{Nir Shavit}.}
  \bibinfo{year}{2005}\natexlab{}.
\newblock \showarticletitle{A lazy concurrent list-based set algorithm}. In
  \bibinfo{booktitle}{\emph{International Conference On Principles Of
  Distributed Systems}}. Springer, \bibinfo{pages}{3--16}.
\newblock


\bibitem[\protect\citeauthoryear{Herlihy, Luchangco, Martin, and Moir}{Herlihy
  et~al\mbox{.}}{2005}]%
        {herlihy2005nonblocking}
\bibfield{author}{\bibinfo{person}{Maurice Herlihy}, \bibinfo{person}{Victor
  Luchangco}, \bibinfo{person}{Paul Martin}, {and} \bibinfo{person}{Mark
  Moir}.} \bibinfo{year}{2005}\natexlab{}.
\newblock \showarticletitle{Nonblocking memory management support for
  dynamic-sized data structures}.
\newblock \bibinfo{journal}{\emph{ACM Transactions on Computer Systems (TOCS)}}
  \bibinfo{volume}{23}, \bibinfo{number}{2} (\bibinfo{year}{2005}),
  \bibinfo{pages}{146--196}.
\newblock


\bibitem[\protect\citeauthoryear{Howley and Jones}{Howley and Jones}{2012}]%
        {howley2012non}
\bibfield{author}{\bibinfo{person}{Shane~V Howley} {and}
  \bibinfo{person}{Jeremy Jones}.} \bibinfo{year}{2012}\natexlab{}.
\newblock \showarticletitle{A non-blocking internal binary search tree}. In
  \bibinfo{booktitle}{\emph{Proceedings of the twenty-fourth annual ACM
  symposium on Parallelism in algorithms and architectures}}.
  \bibinfo{pages}{161--171}.
\newblock


\bibitem[\protect\citeauthoryear{McKenney and Slingwine}{McKenney and
  Slingwine}{1998}]%
        {mckenney1998read}
\bibfield{author}{\bibinfo{person}{Paul~E McKenney} {and}
  \bibinfo{person}{John~D Slingwine}.} \bibinfo{year}{1998}\natexlab{}.
\newblock \showarticletitle{Read-copy update: Using execution history to solve
  concurrency problems}. In \bibinfo{booktitle}{\emph{Parallel and Distributed
  Computing and Systems}}, Vol.~\bibinfo{volume}{509518}.
\newblock


\bibitem[\protect\citeauthoryear{Michael}{Michael}{2004}]%
        {michael2004hazard}
\bibfield{author}{\bibinfo{person}{Maged~M Michael}.}
  \bibinfo{year}{2004}\natexlab{}.
\newblock \showarticletitle{Hazard pointers: Safe memory reclamation for
  lock-free objects}.
\newblock \bibinfo{journal}{\emph{IEEE Transactions on Parallel and Distributed
  Systems}} \bibinfo{volume}{15}, \bibinfo{number}{6} (\bibinfo{year}{2004}),
  \bibinfo{pages}{491--504}.
\newblock


\bibitem[\protect\citeauthoryear{Natarajan and Mittal}{Natarajan and
  Mittal}{2014}]%
        {natarajan2014fast}
\bibfield{author}{\bibinfo{person}{Aravind Natarajan} {and}
  \bibinfo{person}{Neeraj Mittal}.} \bibinfo{year}{2014}\natexlab{}.
\newblock \showarticletitle{Fast concurrent lock-free binary search trees}. In
  \bibinfo{booktitle}{\emph{Proceedings of the 19th ACM SIGPLAN symposium on
  Principles and practice of parallel programming}}. \bibinfo{pages}{317--328}.
\newblock


\bibitem[\protect\citeauthoryear{Nikolaev and Ravindran}{Nikolaev and
  Ravindran}{2019}]%
        {nikolaev2019hyaline}
\bibfield{author}{\bibinfo{person}{Ruslan Nikolaev} {and}
  \bibinfo{person}{Binoy Ravindran}.} \bibinfo{year}{2019}\natexlab{}.
\newblock \showarticletitle{Hyaline: fast and transparent lock-free memory
  reclamation}. In \bibinfo{booktitle}{\emph{Proceedings of the 2019 ACM
  Symposium on Principles of Distributed Computing}}.
  \bibinfo{pages}{419--421}.
\newblock


\bibitem[\protect\citeauthoryear{Nikolaev and Ravindran}{Nikolaev and
  Ravindran}{2020}]%
        {nikolaev2020universal}
\bibfield{author}{\bibinfo{person}{Ruslan Nikolaev} {and}
  \bibinfo{person}{Binoy Ravindran}.} \bibinfo{year}{2020}\natexlab{}.
\newblock \showarticletitle{Universal wait-free memory reclamation}. In
  \bibinfo{booktitle}{\emph{Proceedings of the 25th ACM SIGPLAN Symposium on
  Principles and Practice of Parallel Programming}}. \bibinfo{pages}{130--143}.
\newblock


\bibitem[\protect\citeauthoryear{Prokopec, Bronson, Bagwell, and
  Odersky}{Prokopec et~al\mbox{.}}{2012}]%
        {prokopec2012concurrent}
\bibfield{author}{\bibinfo{person}{Aleksandar Prokopec},
  \bibinfo{person}{Nathan~Grasso Bronson}, \bibinfo{person}{Phil Bagwell},
  {and} \bibinfo{person}{Martin Odersky}.} \bibinfo{year}{2012}\natexlab{}.
\newblock \showarticletitle{Concurrent tries with efficient non-blocking
  snapshots}. In \bibinfo{booktitle}{\emph{Proceedings of the 17th ACM SIGPLAN
  symposium on Principles and Practice of Parallel Programming}}.
  \bibinfo{pages}{151--160}.
\newblock


\bibitem[\protect\citeauthoryear{Ramachandran and Mittal}{Ramachandran and
  Mittal}{2015a}]%
        {ramachandran2015castle}
\bibfield{author}{\bibinfo{person}{Arunmoezhi Ramachandran} {and}
  \bibinfo{person}{Neeraj Mittal}.} \bibinfo{year}{2015}\natexlab{a}.
\newblock \showarticletitle{CASTLE: fast concurrent internal binary search tree
  using edge-based locking}.
\newblock \bibinfo{journal}{\emph{ACM SIGPLAN Notices}} \bibinfo{volume}{50},
  \bibinfo{number}{8} (\bibinfo{year}{2015}), \bibinfo{pages}{281--282}.
\newblock


\bibitem[\protect\citeauthoryear{Ramachandran and Mittal}{Ramachandran and
  Mittal}{2015b}]%
        {ramachandran2015fast}
\bibfield{author}{\bibinfo{person}{Arunmoezhi Ramachandran} {and}
  \bibinfo{person}{Neeraj Mittal}.} \bibinfo{year}{2015}\natexlab{b}.
\newblock \showarticletitle{A fast lock-free internal binary search tree}. In
  \bibinfo{booktitle}{\emph{Proceedings of the 2015 International Conference on
  Distributed Computing and Networking}}. \bibinfo{pages}{1--10}.
\newblock


\bibitem[\protect\citeauthoryear{Ramalhete and Correia}{Ramalhete and
  Correia}{2017}]%
        {ramalhete2017brief}
\bibfield{author}{\bibinfo{person}{Pedro Ramalhete} {and}
  \bibinfo{person}{Andreia Correia}.} \bibinfo{year}{2017}\natexlab{}.
\newblock \showarticletitle{Brief announcement: Hazard eras-non-blocking memory
  reclamation}. In \bibinfo{booktitle}{\emph{Proceedings of the 29th ACM
  Symposium on Parallelism in Algorithms and Architectures}}.
  \bibinfo{pages}{367--369}.
\newblock


\bibitem[\protect\citeauthoryear{Shafiei}{Shafiei}{2013}]%
        {shafiei2013non}
\bibfield{author}{\bibinfo{person}{Niloufar Shafiei}.}
  \bibinfo{year}{2013}\natexlab{}.
\newblock \showarticletitle{Non-blocking Patricia tries with replace
  operations}. In \bibinfo{booktitle}{\emph{2013 IEEE 33rd International
  Conference on Distributed Computing Systems}}. IEEE,
  \bibinfo{pages}{216--225}.
\newblock


\bibitem[\protect\citeauthoryear{Timnat and Petrank}{Timnat and
  Petrank}{2014}]%
        {timnat2014practical}
\bibfield{author}{\bibinfo{person}{Shahar Timnat} {and} \bibinfo{person}{Erez
  Petrank}.} \bibinfo{year}{2014}\natexlab{}.
\newblock \showarticletitle{A practical wait-free simulation for lock-free data
  structures}.
\newblock \bibinfo{journal}{\emph{ACM SIGPLAN Notices}} \bibinfo{volume}{49},
  \bibinfo{number}{8} (\bibinfo{year}{2014}), \bibinfo{pages}{357--368}.
\newblock


\bibitem[\protect\citeauthoryear{Wen, Izraelevitz, Cai, Beadle, and Scott}{Wen
  et~al\mbox{.}}{2018}]%
        {wen2018interval}
\bibfield{author}{\bibinfo{person}{Haosen Wen}, \bibinfo{person}{Joseph
  Izraelevitz}, \bibinfo{person}{Wentao Cai}, \bibinfo{person}{H~Alan Beadle},
  {and} \bibinfo{person}{Michael~L Scott}.} \bibinfo{year}{2018}\natexlab{}.
\newblock \showarticletitle{Interval-based memory reclamation}.
\newblock \bibinfo{journal}{\emph{ACM SIGPLAN Notices}} \bibinfo{volume}{53},
  \bibinfo{number}{1} (\bibinfo{year}{2018}), \bibinfo{pages}{1--13}.
\newblock


\end{thebibliography}

\section{Artifact Description}
\label{sec:aedescp}
This section provides a step by step guide to run our artifact (nbr\_setbench) in a docker container. Other ways to setup a machine to use our artifact are provided in the \textit{readme} file at the URL: \url{https://doi.org/10.5281/zenodo.4409185}.

To better reproduce the results described in our paper we recommend to run the nbr\_setbench on a NUMA machine with atleast 2 NUMA nodes having a recent Linux distro (we used Ubuntu 18.04 or 20.04) with 188GB RAM and recent Docker installation (we used version 19.03.6, build 369ce74a3c). 
\\
\\
\myparagraph{Steps to load and run the provided Docker image:}
Sudo permission may be required to execute the following instructions. 

\begin{enumerate}
    \item Install the latest version of Docker on your system. We tested the artifact with the Docker version 19.03.6, build 369ce74a3c. Instructions to install Docker may be found at\\ \url{https://docs.docker.com/engine/install/ubuntu}.
\begin{verbatim}
$ docker -v
\end{verbatim}

    \item Download the artifact from Zenodo at URL:\\
    \url{https://doi.org/10.5281/zenodo.4409185}.

    \item Extract the downloaded folder and move to \\
    \textit{nbr\_setbench/} directory using $cd$ command.
    \item Find docker image named \textit{nbr\_docker.tar.gz}\\
    in nbr\_setbench/ directory. And load the downloaded docker image with the following command:
\begin{verbatim}
 $ sudo docker load -i nbr_docker.tar.gz
\end{verbatim}
\item 
Verify that image was loaded:
\begin{verbatim}
 $ sudo docker images
\end{verbatim}
\item Start a docker container from the loaded image:
\begin{verbatim}
 $ sudo docker run --name nbr -i -t \ 
 --privileged nbr_setbench /bin/bash
\end{verbatim}
\item Invoke $ls$ to see several files/folders of the artifact: Dockerfile, README.md, common, ds, install.sh, lib, microbench, nbr\_experiments, tools.
\end{enumerate}

\myparagraph{Steps to run the experiments:} To compile, run and see results of the experiment follow these steps:

\textbf{Input}: Inputs to the experiment can be configured in corresponding input files at: 

\textit{/nbr\_setbench/nbr\_experiments/inputs/}

\textbf{Output}: Generated figures can be found in directory:

\textit{/nbr\_setbench/nbr\_experiments/plots/generated\_plots/}

\begin{enumerate}
    \item Assuming you are currently in $nbr\_setbench$, execute the following command:
\begin{verbatim}
 $ cd nbr_experiments/
\end{verbatim} 
\item Run the following command to generate plots for \\throughput evaluation:
\begin{verbatim}
 $ ./run.sh
\end{verbatim} 
\item Run the following command to generate plots for memory usage evaluation:
\begin{verbatim}
 $ ./run_memusage.sh
\end{verbatim} 
\end{enumerate}

After the above scripts finish executing DO NOT exit the terminal as we would need to copy the generated
figures on the host machine to be able to see them.

\myparagraph{Steps to visualize the plots:} 
Resultant figures could be found in 
\textit{nbr\_experiments/plots/generated\_plots}.

To visualize the generated figures on your host machine copy the plots from the docker container to your host system by following these steps:

\begin{enumerate}
    \item Verify the name of the docker container. Use the following command which would give us the name of the loaded docker container under NAMES column which is 'nbr'.
\begin{verbatim}
$ sudo docker container ls
\end{verbatim} 
    \item Open a new terminal on the same machine. Move to any directory where you would want the generated plots to be copied (use cd). And execute the following command to copy the generated plots from the \textit{nbr\_experiments/plots/generated\_plots} folder to your current directory.
\begin{verbatim}
$ sudo docker cp nbr:/nbr_setbench/ \ 
 nbr_experiments/plots/generated_plots/ .
\end{verbatim} 
\end{enumerate}

Now you can analyse the generated plots.
Each plot follows a naming convention: 
\begin{enumerate}
\item throughput-[data structure name]-[number of inserts]-[number of deletes].png. For
example, a plot showing throughput of DGT with 50\% inserts and 50\% deletes is named as: throughput-guerraoui\_ext\_bst\_ticket-i50-d50.png.

\item Similarly the plot for peak memory usage experiments follows a naming convention:\\ mem\_usage-[data structure name]-[number of inserts]-[number of deletes].png. For example, a plot showing mem\_usage of DGT with 50\% inserts and 50\% deletes is named as: mem\_usage-guerraoui\_ext\_bst\_ticket-i50-
d50.png.
\end{enumerate}

\appendix
\section{Correctness}
\label{sec:apncor}
\begin{assumption}
\label{asm:sigg}
If \xspace \tid{i} send a signal to \tid{j}, then \tid{j} is guaranteed to receive it and execute a signal handler before dereferencing any reference field of a record. 
\end{assumption}

\begin{property}
\label{prop:rdp}
A thread \tid{j} in \rdp
\begin{enumerate}[label=\ref{prop:rdp}.\arabic*]
    \item \label{prop:rdp1} {upon receiving a signal executes a signal handler and gets neutralized.}
    \item \label{prop:rdp2} {could dereference the reference field of a record to discover new records.}
\end{enumerate}
\end{property}

\begin{property}
\label{prop:wtp}
A thread \tid{j} in \wtp
\begin{enumerate}[label=\ref{prop:wtp}.\arabic*]
    \item \label{prop:wtp1} {protects all records used in \wtp and upon receiving a signal executes a signal handler.}
    \item \label{prop:wtp2} {could not dereference the reference field of a record to discover unprotected records.}
\end{enumerate}
\end{property}

\begin{property}
Every \rl thread \tid{r} does the following in order:
\label{prop:rl}
\begin{enumerate}[label=\ref{prop:rl}.\arabic*]
    \item \label{prop:rl1} {sends signal to all participating threads.}
    \item \label{prop:rl2} {scans all protected records of each participating thread \tid{j}.}
    \item \label{prop:rl3} {reclaims records which are not protected.}
\end{enumerate}
\end{property}

\begin{lemma}[NBR is Safe]
\label{lem:nbr}
No \rl thread in \nbr would reclaim a record that is not safe.
\end{lemma}
\begin{proof}
$Wlog$, assume, the statement is false. Implying, there exists a \rl \tid{r} which reclaims an unsafe record $rec$. This could occur only in two ways: (1) a \rd dereferences a reference to $rec$ in limboBag of \tid{r}, or (2) a \wt dereferences $rec$ which it did not protect. 

We show that (1) is false:
Due to \propref{rl1} \tid{r} must have sent signal to \rd before reclaiming $rec$. From, \asmref{sigg} and \propref{rdp1}, \rd  is guaranteed to execute signal handler as next step in its execution, hence will be neutralized relinquishing its private reference to $rec$, if any.

Next we show that (2) is also false:
Again, Due to \propref{rl1} \tid{r} must have sent signal to \wt before reclaiming $rec$. From \propref{wtp} \wt must have protected $rec$ and any reference which it could access through reference field of $rec$.

Thus, (1) and (2) are false, meaning $rec$ is safe  which contradicts our assumption that $rec$ is unsafe.
\end{proof}

\begin{lemma}[\nbrp is safe]
No \rl thread in \nbrp would reclaim a record that is not safe.
\end{lemma}
\begin{proof}
\nbrp has two kind of threads that reclaim (1) threads at \lw and (2) threads at \hw. Reclamation at \hw is similar to reclamation in \nbr. Since \nbr is safe(refer \lemref{nbr}),  reclamation at \hw in \nbrp is safe. Therefore, our task reduces to prove that reclamation at \lw in \nbrp is safe.

Assume, \textbf{(A)} reclamation by threads at \lw is not safe. Let \tid{{lw}} be such a thread. Then following case must be true. \textbf{(C)} Their is a record, say $rec$, in limboBag of \tid{{lw}} to which another thread, say \tid{r}, holds an unsafe private reference.

But, in \nbrp, \tid{{lw}} reclaims only upto the \emph{bookmarkTail}, which depicts a time, say $t$, at which \tid{{lw}} entered \lw. And, \tid{{lw}} decides to reclaim only at later time $t'$, such that $t<t'$, which corresponds to a \emph{relax grace period} (\rgp). By definition, \rgp implies that any record unlinked before it would be safe as all threads would have discarded all unreserved references. Thus, no thread could hold any private reference to $rec$ which was unlinked before $t$ after time $t'$, since $t<t'$. Implying, \textbf{(C)} is false. Consequently, our assumption \textbf{(A)} is false. Since, \textbf{(A)} cannot be both true and false. Therefore, by contradiction, it is established that \nbrp is safe. 
\end{proof}

\begin{lemma}[Both \nbr and \nbrp are robust] Number of records that could stay un-reclaimed per thread are bounded.
\end{lemma}
\begin{proof}
Assuming data-structure using \nbr is correct. Thus, a record is passed as an argument to \textit{retire} only once. Consequently, an unlinked record would be present in exactly one thread's limboBag.
Say, $k$ is number of records a thread could protect per operation, $p$ is number of processes, and $n$ is maximum limboBag size at which a thread decides to reclaim. Usually $p << n$ and $k << n$.

Let, \tid{r} be a \rl and \tid{j} be an arbitrary thread. Now, if \tid{j} is delayed or crashed, it could only reserve atmost $k$ records of \tid{r}'s limboBag. Thus, in worst case a single thread could prevent only $k$ records from being reclaimed. Inducting on number of threads, all $p-1$ threads could prevent only $k(p-1)$ records in any thread's limbo bag from being reclaimed.

Since, $p<<n$, $k(p-1) << n$. Thus, number of records which could stay un-reclaimed per thread would be $\theta(kp)$.   
\end{proof}

\begin{corollary}
In worst case, after a reclamation event $\Theta(p)$ records could stay un-reclaimed.
\end{corollary}

\section{Applicability}
\label{sec:apnappl}
In this section we will reason about why HP and \nbr are applicable to certain data structure and not to several others as enumerated in \tabref{appltab}.

In order to safely apply \nbr to a concurrent data structure either 1) the data structure should offer an operation consisting of a single read and single write phase or 2) if it has multiple read-write phases then each read-write phase pair should appear as if it is a separate operation. To be more specific, the latter requirement says that each read phase of the data structure should restart from the \emph{root}. Additionally, one should be able to reserve all pointers that would be needed in a \wtp beforehand. This is to ensure correct handshakes between \rd{s}, \rl{s} and \wt{s}. We formally state these requirements as follows:

\begin{requirement}
\label{req:reqnbr}
Each \rdp that occurs after a \wtp in a data structure should restart from the root.
\end{requirement}
Intuitively, this implies that each read/write phase pair should appear as if it is a separate operation such that pointers to shared memory locations acquired in the previous operation are discarded. 
\begin{requirement}
\label{req:reqnbr2}
All shared pointers that could be accessed in a \wtp should be reserved before entering the \wtp.
\end{requirement}
Intuitively, this implies that once a \wtp starts no new (unreserved) references should be accessed.

In the case of \textbf{HP}, reasoning about the requirements desirable for their safe application to a data structure is more subtle. Conceptually, accessing shared records (memory locations) in data structure operations using HP is a three step process where pointer to records are protected in a hand over hand manner.
Assume a thread is traversing from $rec1$ to $rec2$ by following a pointer to $rec2$ which was read from the field of $rec1$. Just after reading the field of $rec1$ the thread has local pointers to both of records.
Without loss of generality, assume that $rec1$ is already protected and the thread is attempting to protect $rec2$. In order to successfully protect $rec2$ the thread should follow these steps:
\begin{enumerate}
    \item \textbf{Announcement step:} announce a pointer to $rec2$ at a \emph{swmr} memory location so that it becomes visible in a timely manner to all other participating threads. This usually requires using \func{mfence} or \func{xchg}.
    \item \textbf{Reachability validation step:} between the interval when a thread gets a local pointer to $rec2$ and before the thread announces the pointer to it use-after-free access may happen if $rec2$ is freed. 
    Therefore, in order to prevent use-after-free accesses, it is required to verify that $rec2$ is still reachable from the root. This is usually accomplished by checking that $rec2$ is reachable from $rec1$ and $rec1$ is not marked. 
    A more detailed discussion can be found in Brown's paper which proposed a fast EBR based SMR algorithm, DEBRA\cite{brown2015reclaiming}.
    \item \textbf{Acquisition step:} If the \textit{Reachability validation step} is successful then thread is said to have \textit{acquired} a HP on $rec2$ and any subsequent dereference of $rec2$ will be safe. Otherwise, the thread has to retry protecting the $rec2$.
    This is a conceptual step that depends on the outcome of the reachability validation step.
 
\end{enumerate}

We formally state these requirements of safely applying HPs to a data structure as follows:
\begin{requirement}
\label{req:reqhp}
A thread that reads a shared record should \textbf{acquire} a HP on the record before using it.
\end{requirement}

Thus, in order to safely use HPs with a data structure, ~\reqref{reqhp} should be satisfied so that the safety and progress guarantees of the original data structure are not violated.

Keeping these requirements in mind we will first reason about the applicability of HP to the data structures in \tabref{appltab} followed by the applicability of \nbr. 

\subsection{Applicability of HPs}
In LL05\cite{heller2005lazy} searches are wait-free and updates use an optimistic locking protocol. If HPs were applied to LL05 it would be possible for a thread to repeatedly fail to acquire a HP (due to \reqref{reqhp}) on a node that is marked but not yet unlinked. This breaks the wait-freedom for searches in the original data structure as is also explained in \cite{brown2015reclaiming}. Therefore, HPs can not be used with LL05.

On the other hand, HL01 and HM04 do not allow traversals over marked nodes. Thus, the issue of repeated failure to acquire a HP does not occur in HL01 or HM04. Meaning that HPs could be applied safely as shown in \cite{michael2004hazard}. 

DGT2015\cite{david2015asynchronized}, on the other hand, traverses the nodes in a sync-free manner (similar to the lazylist\cite{heller2005lazy}) and uses version numbers and ticket locks to optimistically execute updates. However, since DGT15 does not use marking, it is not clear how a thread could perform reachability validation.

In HJ12, reachability validation could fail indefinitely if a thread that marks a node crashes before unlinking it. More specifically, threads that try to help this crashed thread to finish unlinking the node would need to acquire a HP on the marked node, which would be impossible. This could block all threads from making progress. Such issues were described in \cite{brown2015reclaiming}. Similarly, in HL17~\cite{he2017deletion} and BPA20~\cite{brown2020non} it is not clear how a thread trying to acquire a HP on a node can verify that the node is still reachable from the root. This makes it complex to apply HPs to such data structures without significant modifications which may break progress guarantees.

BCCO10~\cite{bronson2010practical} is a partially external relaxed AVL tree that employs version-numbers and optimistic concurrency control (OCC) techniques to avoid locking nodes in searches as much as possible. 
It uses lazy deletion in the sense that to delete a node with two children it converts the node to a routing node and this routing node is unlinked lazily during subsequent rebalancing. 
Thus, it appears that HPs can be used with BCCO10 as one can leverage version-based validation to know when a node is definitely reachable. If such a validation fails one can simply restart from the root. Note, this doesn't theoretically impact the progress guarantee of searches (unlike LL05 or DGT15) as they are already blocking because of hand over hand optimistic validation. However, in practice, HPs would necessitate restarts from the root when validation fails, whereas BCCO10 nominally attempts to continue traversal from the parent node in this case.

More recent partially external AVL tree DVY14\cite{drachsler2014practical} is based on BCCO10, but it does not use version numbers, so a search has no way to tell whether a node is currently in the tree. Thus, it is not clear how one could use HPs with DVY14.

\subsection{Applicability of NBR}
Reasoning about applicability of \nbr is comparatively easy as one just needs to confirm whether every \rdp restarts from root and if it is possible to reserve all records before entering a \wtp, \reqref{reqnbr} and ~\ref{req:reqnbr2}, respectively. Or in other words, each thread should restart from root after doing updates (possibly to help other threads) in search phase and no new pointers to shared records should be accessed in a \wtp.

HL01\cite{harris2001pragmatic}, as explained in \secref{appl}, can be used with \nbr since its operations have multiple \nbr read-write phases where each read phase restarts from the root ( satisfying \reqref{reqnbr}) and at the end of each \rdp the records which can be accessed inside a subsequent \wtp are known beforehand (fulfilling \reqref{reqnbr2}).

HM04\cite{michael2004hazard} can not directly be used with \nbr since its operations have multiple read-write phases where \rdp phases do not restart from the root. Each \rdp after a preceding auxiliary update (\wtp) resumes from the last reachable node (\emph{pred}) found in the list (violating \reqref{reqnbr}). However, as we discussed in \secref{appl}, the list can be safely modified such that each \rdp restarts from the root and thus \nbr can be used with HM04 with modification.

\nbr can be applied to EFRB10\cite{ellen2010non} as its update operations have lock-less searches that either end with the intended update or \textit{helping} another conflicting update to complete (if the intended update fails). Helping is realized by using \textit{info records} that are similar to descriptors commonly used in lock-free data structures. 
In the former case, \nbr can simply reserve the nodes returned by the search and execute the update within a \wtp (\reqref{reqnbr2}). In the latter case, after helping another update complete, a thread restarts from the root (\reqref{reqnbr}). Also, before entering the \wtp, to help another conflicting update, the thread can simply reserve all pointers in fields of Info records (similar to descriptors commonly used in lock-free data structures) and then do the update (\reqref{reqnbr2}). However, \nbr can not be applied to EFRB14\cite{ellen2014amortized} as after helping, a thread restarts from nearby ancestors and not the root, which is a violation of \reqref{reqnbr}.

HJ12\cite{howley2012non}, the first lock-free internal BST is loosely based on EFRB10. Here, the searches are required to do a helping update to avoid missing a node in the tree because some concurrent update may have moved the node up in its search path. Specifically, this occurs when a search of a node overlaps with a two-child delete operation. However, in HJ12, after each helping update, searches restart from the root (\reqref{reqnbr}) and all records required to do the helping update can be known beforehand through descriptors (\reqref{reqnbr2}). Thus, HJ12 applies to \nbr.

Similarly, data structures designed using Brown, Ellen and Ruppert's tree update template \cite{brown2014general}, for example lock-free chromatic tree\cite{brown2014general}, ABTree, AVL trees\cite{brown2017techniques} and HL17\cite{he2017deletion} could be used with \nbr.
In the template, operations do sync-free searches to find target node(s) (which is similar to standard \rdp), then check whether they need to help (auxiliary updates, similar to entering \wtp during searches), and after the helping, operations restart from the root. 
Since \textit{helping} involves descriptors which contain pointers to all nodes that would be required to execute the auxiliary update, \nbr could reserve all node pointers in the descriptor and enter \wtp for the auxiliary update and then start subsequent search (\rdp) from the root. 
However, a more recently proposed external interpolation search tree cannot be used with \nbr details of which are explained in \secref{appl}.

lock-free patricia trie S13\cite{shafiei2013non} too follows a pattern in which searches are sync-free. And updates may do auxiliary helping using descriptors (thus, all pointers to be accessed during updates can be known beforehand.) and then restart from the root. As a consequence, \nbr applies to S13 since both \reqref{reqnbr} and ~\ref{req:reqnbr2} are satisfied.  

DGT15\cite{david2015asynchronized} has sync-free searches followed by locking one node for inserts, two nodes for deletes, then modifying the nodes. It is similar to the single \rdp followed by a single \wtp design pattern of LL05\cite{heller2005lazy}. Thus, \nbr applies to DGT15.

Unbalanced external BST of Natarajan and Mittal (NM14) \cite{natarajan2014fast} too has a pattern where each operation starts with a sync-free search that returns a \textit{seek record} object followed by updates.
During the updates, threads may possibly help by accessing nodes only pointed by the \textit{seek record} object and then subsequently restart from the root. It is along the lines of \nbr{'s} \reqref{reqnbr} and ~\ref{req:reqnbr2}.
Deletion consists of two modes: injection and cleanup. Both involve data structure modification, therefore, both should be done in \nbr{'s} \wtp. Additionally, it is possible that injection mode may succeed, and subsequent cleanup may fail, in that case, the operation is required to start from the root. In sum, NM14 satisfies the requirements making it suitable to be used with \nbr. 

The unbalanced external BST in DVY14a\cite{drachsler2014practical} does a sync-free search and then executes updates using locks. But inside the update phase, it may obtain pointers to new nodes, for example, successor, children or parent of a node. This violates the \reqref{reqnbr2} of \nbr.
As a consequence, it seems that \nbr can not be applied to DVY14a.
However, it is possible to make it work with \nbr. For that one must modify DVY14a to perform all of the reads of new node pointers (including those in the update phase) before the first lock is acquired, then validate after lock acquisitions that the values of those reads have not changed (and restart if validation fails).
Note, \nbr would not work with their balanced variant of this tree (DVY14b\cite{drachsler2014practical}), which does bottom-up rebalancing without restarting from the root between rotations.

BCCO10~\cite{bronson2010practical} uses optimistic concurrency control (OCC) techniques to avoid locking nodes in searches as much as possible but occasionally locks and immediately unlocks a node as part of traversal, which suggests NBR cannot be used. 
However, as described in~\cite{arbel2018getting}, this locking is an optional part of the BCCO algorithm (intended to improve fairness under heavy contention) and can simply be removed.
Unfortunately, this algorithm performs recursive bottom-up rebalancing without restarting from the root between rotations. To use \nbr, one would need to restart from the root after each rotation, which would be a substantial algorithmic change.

Unbalanced internal BST of Ramachandran and Mittal (RM15)\cite{ramachandran2015fast} ensures that a \emph{search} operation for a key $k$ does not miss $k$ if it is moved upwards by having the search save a pointer to a node they term as \textit{anchor node}. If the \emph{search} does not find $k$, it checks the \textit{anchor node} to see if $k$ was moved.
Whereas, an insert operation, first does a sync-free search, then does a CAS on an appropriate child, and returns if the CAS succeeds. If it fails, the insert re-reads the pointer it failed to change to check whether the pointer points to a descriptor (which might need help). 
If the pointer is a descriptor, then the thread doing the insert helps it and subsequently restarts from the root.
The fact that a thread re-reads the pointer after performing a CAS, and helps whatever descriptor it finds, is a problem for \nbr.
This would mean dereferencing a pointer obtained after entering the write phase without restarting from the root--- violation of the \reqref{reqnbr}. 
To use \nbr, in order to help safely, we must reserve both children of a node before we do the aforementioned CAS, and if either child is actually a descriptor, we must reserve all pointers in the descriptor (so we can safely dereference them if needed during helping). At this point, if we see a descriptor, we might as well help it before attempting the CAS (if the CAS would be doomed to fail anyway). Unfortunately, this changes the progress argument (although we don't think it actually changes the progress property).
However, deletes in RM15 are the real problem for applying \nbr because the two child deletion case requires dereferencing new pointers without restarting from the root. Modifying this algorithm to work with \nbr would require sweeping changes. Thus, \nbr cannot be applied to RM15.

\section{Additional Experiments}
\label{sec:apnexp}
\begin{figure*}[htbp]
     \begin{minipage}{\textwidth}
        \begin{subfigure}{\textwidth}
            \includegraphics[width=0.33\linewidth, height=5cm, keepaspectratio]{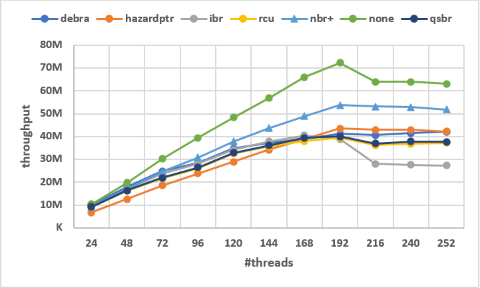}\hfill
            \includegraphics[width=0.33\linewidth, height=5cm, keepaspectratio]{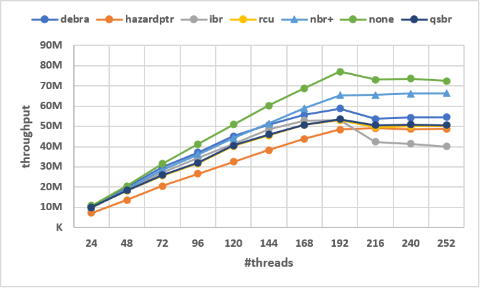}\hfill
            \includegraphics[width=0.33\linewidth, height=5cm, keepaspectratio]{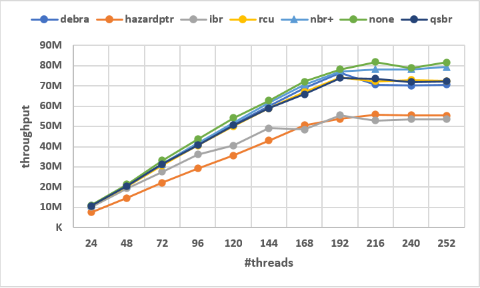}\hfill
            \caption{DGT-tree. Left: 50i-50d. Middle: 25i-25d. Right: 5i-5d. Max size:20M.}
            \label{fig:apntree20M}
        \end{subfigure}
        \begin{subfigure}{\textwidth}
            \includegraphics[width=0.33\linewidth, height=5cm, keepaspectratio]{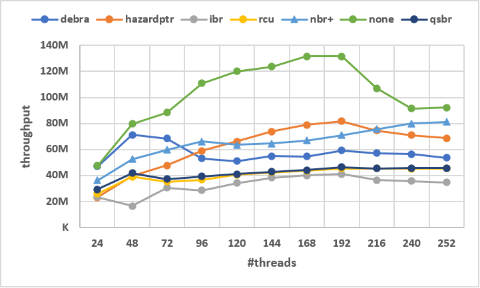}\hfill
            \includegraphics[width=0.33\linewidth, height=5cm, keepaspectratio]{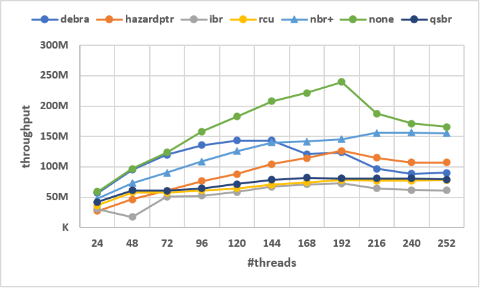}\hfill
            \includegraphics[width=0.33\linewidth, height=5cm, keepaspectratio]{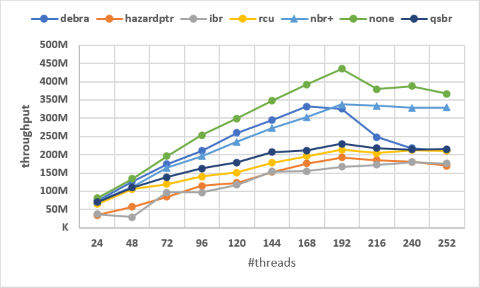}\hfill
            \caption{DGT-tree. Left: 50i-50d. Middle: 25i-25d. Right: 5i-5d. Max size:20K.}
            \label{fig:apntree20K}
        \end{subfigure}
     \end{minipage}
    \caption{E1: Evaluation of throughput across different DGT sizes. Y axis: throughput in million operations per second. X axis: \#threads.}
    \label{fig:apntree}
\end{figure*}

\begin{figure*}[h]
     \begin{minipage}{\textwidth}
        \begin{subfigure}{\textwidth}
            \includegraphics[width=0.33\linewidth, height=5cm, keepaspectratio]{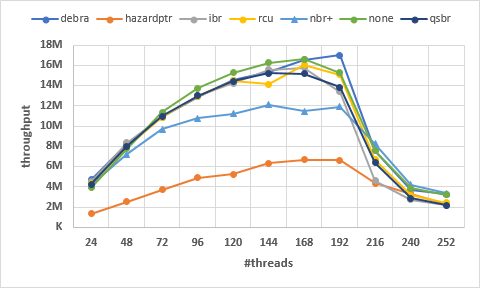}\hfill
            \includegraphics[width=0.33\linewidth, height=5cm, keepaspectratio]{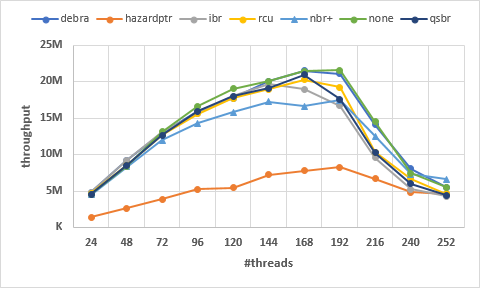}\hfill
            \includegraphics[width=0.33\linewidth, height=5cm, keepaspectratio]{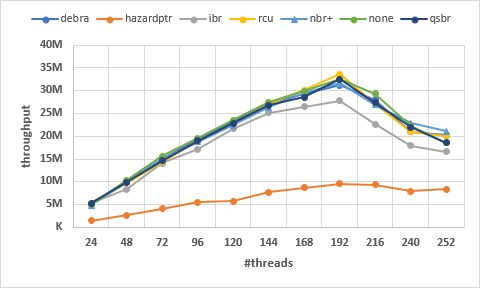}\hfill
            \caption{lazy list. Left: 50i-50d. Middle: 25i-25d. Right: 5i-5d. Max size:2K.}
            \label{fig:apnlist2K}
        \end{subfigure}
        \begin{subfigure}{\textwidth}
            \includegraphics[width=0.33\linewidth, height=5cm, keepaspectratio]{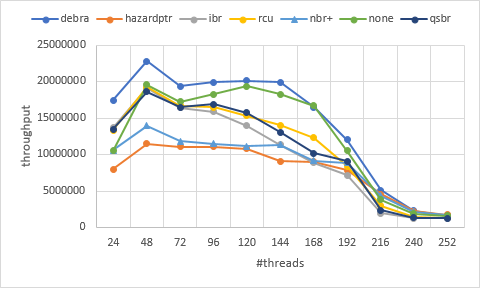}\hfill
            \includegraphics[width=0.33\linewidth, height=5cm, keepaspectratio]{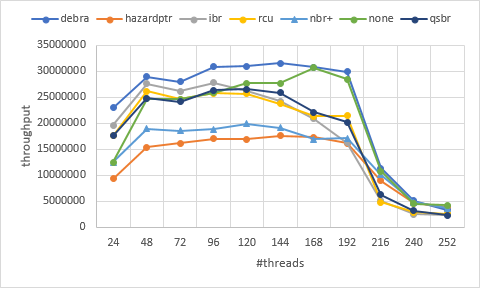}\hfill
            \includegraphics[width=0.33\linewidth, height=5cm, keepaspectratio]{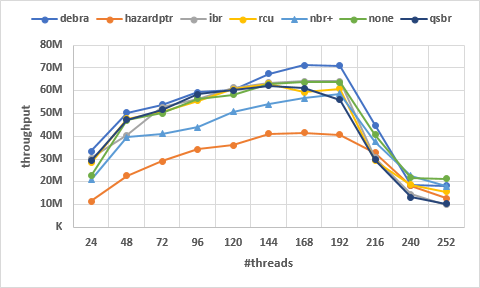}\hfill
            \caption{lazy list. Left: 50i-50d. Middle: 25i-25d. Right: 5i-5d. Max size:200.}
            \label{fig:apnlist200}
        \end{subfigure}
     \end{minipage}
    \caption{E1: Evaluation of throughput across different lazylist (LL05) sizes. Y axis: throughput in million operations per second. X axis: \#threads.}
    \label{fig:apnlist}
\end{figure*}

This section contains additional results for different data structure sizes.
\figref{apntree20M} and \figref{apntree20K} show throughput of DGT for the size of 20M and 20K, respectively. As can be observed, \nbrp is faster than other techniques across the size of 20K (high contention) and 20M (low contention).

For the lazy list (\figref{apnlist2K} and \figref{apnlist200}) in extremely high contention (size 200 and update-intensive workload) \nbrp is slower than the EBR based variants but still comparable to HP. The degradation in throughput could be attributed to the overhead of the signal apparatus \textendash frequent neutralizing signals, \func{siglongjmp} and \func{sigsetjmp} which becomes prominent due to high number of updates and small list size relative to other SMR algorithms.

\begin{figure*}[h]
     \begin{minipage}{\textwidth}
        \begin{subfigure}{\textwidth}
            \includegraphics[width=0.33\linewidth, height=5cm, keepaspectratio]{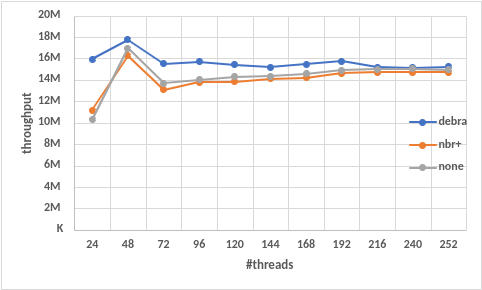}\hfill
            \includegraphics[width=0.33\linewidth, height=5cm, keepaspectratio]{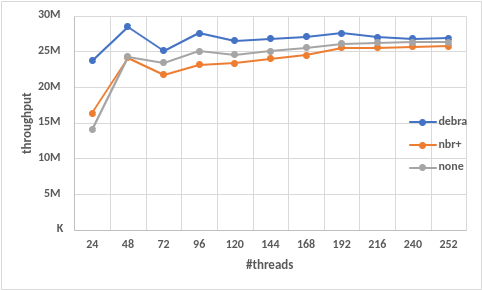}\hfill
            \includegraphics[width=0.33\linewidth, height=5cm, keepaspectratio]{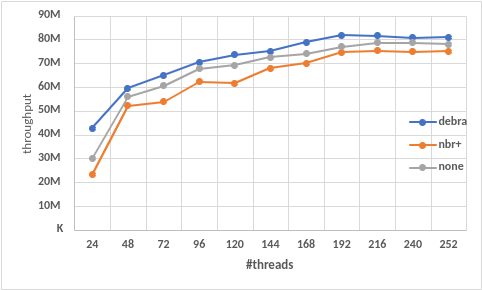}\hfill
            \caption{Harris list. Left: 50i-50d. Middle: 25i-25d. Right: 5i-5d. Max size:200.}
            \label{fig:apnhlist200}
        \end{subfigure}     
        \begin{subfigure}{\textwidth}
            \includegraphics[width=0.33\linewidth, height=5cm, keepaspectratio]{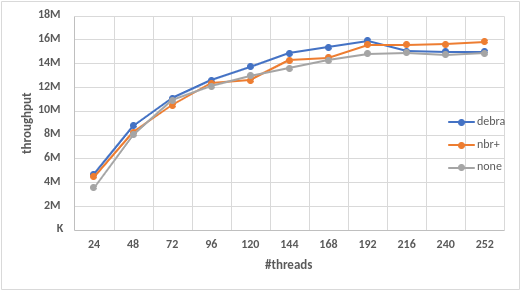}\hfill
            \includegraphics[width=0.33\linewidth, height=5cm, keepaspectratio]{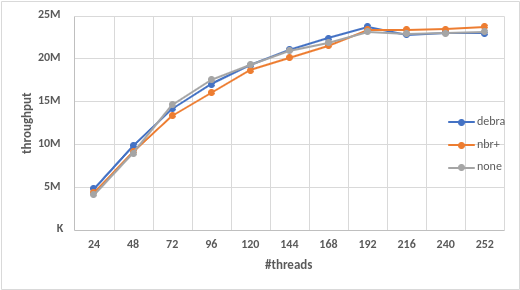}\hfill
            \includegraphics[width=0.33\linewidth, height=5cm, keepaspectratio]{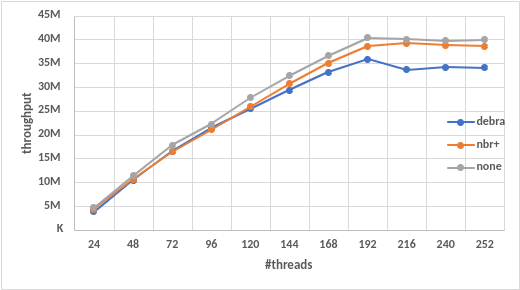}\hfill
            \caption{Harris list. Left: 50i-50d. Middle: 25i-25d. Right: 5i-5d. Max size:2K.}
            \label{fig:apnhlist2K}
        \end{subfigure}
        \begin{subfigure}{\textwidth}
            \includegraphics[width=0.33\linewidth, height=5cm, keepaspectratio]{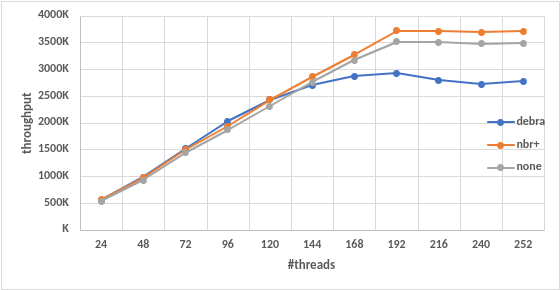}\hfill
            \includegraphics[width=0.33\linewidth, height=5cm, keepaspectratio]{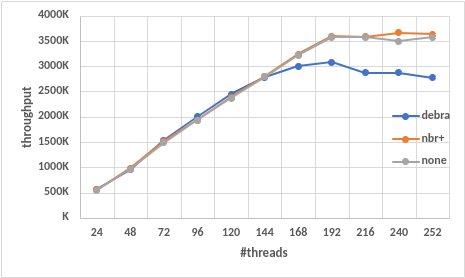}\hfill
            \includegraphics[width=0.33\linewidth, height=5cm, keepaspectratio]{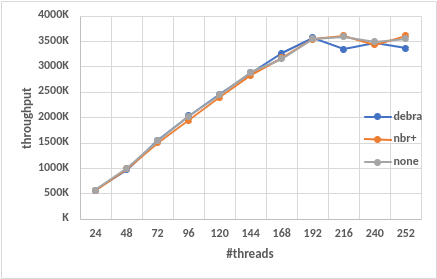}\hfill
            \caption{Harris list. Left: 50i-50d. Middle: 25i-25d. Right: 5i-5d. Max size:20K.}
            \label{fig:apnhlist20K}
        \end{subfigure}
     \end{minipage}
    \caption{E3: Evaluation of throughput across different Harris list(HL01) sizes. Y axis: throughput in million operations per second. X axis: \#threads.}
    \label{fig:apnhlist}
\end{figure*}

\begin{figure*}[t]
     \begin{minipage}{\textwidth}
        \begin{subfigure}{\textwidth}
            \includegraphics[width=0.33\linewidth, height=5cm, keepaspectratio]{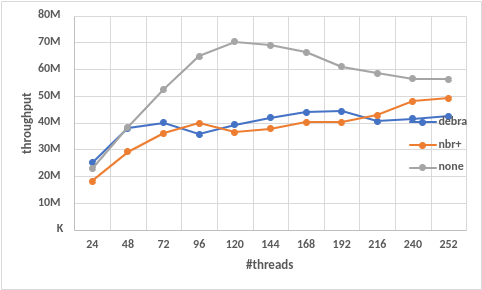}\hfill
            \includegraphics[width=0.33\linewidth, height=5cm, keepaspectratio]{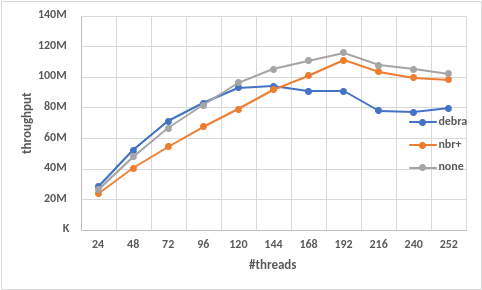}\hfill
            \includegraphics[width=0.33\linewidth, height=5cm, keepaspectratio]{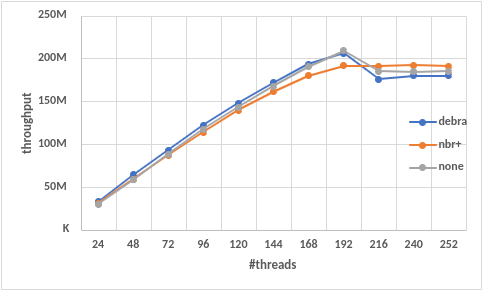}\hfill
            \caption{ABTree. Left: 50i-50d. Middle: 25i-25d. Right: 5i-5d. Max size:20M.}
            \label{fig:apnabtree20M}
        \end{subfigure}
        \begin{subfigure}{\textwidth}
            \includegraphics[width=0.33\linewidth, height=5cm, keepaspectratio]{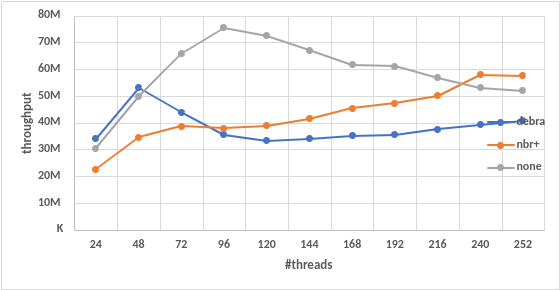}\hfill
            \includegraphics[width=0.33\linewidth, height=5cm, keepaspectratio]{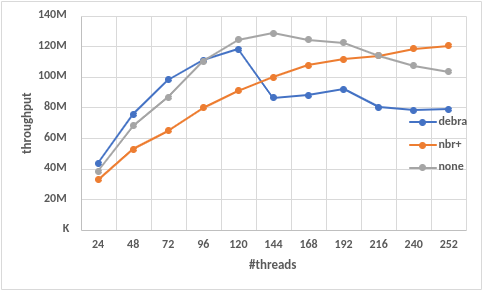}\hfill
            \includegraphics[width=0.33\linewidth, height=5cm, keepaspectratio]{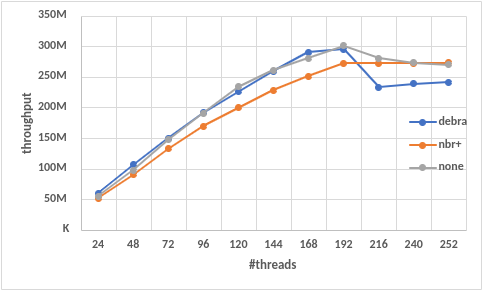}\hfill
            \caption{ABTree. Left: 50i-50d. Middle: 25i-25d. Right: 5i-5d. Max size:2M.}
            \label{fig:apnabtree2M}
        \end{subfigure}
     \end{minipage}
    \caption{E3: Evaluation of throughput across different ABTree sizes. Y axis: throughput in million operations per second. X axis: \#threads.}
    \label{fig:apnabtree}
\end{figure*}

\figref{apnabtree} and ~\ref{fig:apnhlist} are the extension of the third experiment (E3).





\end{document}